\documentclass[twocolumn,pra,groupedaddress,superscriptaddress,nofootinbib]{revtex4}
\usepackage{graphicx,amsmath,amssymb,txfonts,float}
\usepackage{subfigure,hyperref,bbm,times}
\usepackage[T1]{fontenc}
\usepackage{braket}
\usepackage{epsfig}
\usepackage{color}
\usepackage{graphicx}
\usepackage{dcolumn}
\usepackage{bm}

\providecommand{\openone}{\leavevmode\hbox{\small1\kern-3.8pt\normalsize1}}

\usepackage{soul}

\renewcommand{\tilde}{~}
\newcommand{\id}{1\!\!1}
\newcommand{\kluru}{\ket{L\uparrow,R\uparrow}}
\newcommand{\klurd}{\ket{L\uparrow,R\downarrow}}
\newcommand{\kldru}{\ket{L\downarrow,R\uparrow}}
\newcommand{\kldrd}{\ket{L\downarrow,R\downarrow}}

\newcommand{\kaubu}{\ket{A\uparrow,B\uparrow}}
\newcommand{\kaubd}{\ket{A\uparrow,B\downarrow}}
\newcommand{\kadbu}{\ket{A\downarrow,B\uparrow}}
\newcommand{\kadbd}{\ket{A\downarrow,B\downarrow}}

\newcommand{\rlr}{\rho_{\textrm{LR}}}
\newcommand{\rab}{\rho_{\textrm{AB}}}
\newcommand{\rtilde}{\widetilde{\rho}}
\newcommand{\td}{t_\textrm{D}}
\newcommand{\rd}{\rho_\textrm{D}}
\newcommand{\ku}{\ket{\uparrow}}
\newcommand{\bu}{\bra{\uparrow}}
\newcommand{\kd}{\ket{\downarrow}}
\newcommand{\bd}{\bra{\downarrow}}

\newcommand{\kop}{\ket{1_+}}
\newcommand{\kom}{\ket{1_-}}
\newcommand{\ktp}{\ket{2_+}}
\newcommand{\ktm}{\ket{2_-}}
\newcommand{\bop}{\bra{1_+}}
\newcommand{\bom}{\bra{1_-}}

\newcommand{\ab}{\textrm{AB}}
\newcommand{\dkom}{\ket{\bar{1}_{-}}}
\newcommand{\dkop}{\ket{\bar{1}_{+}}}
\newcommand{\dktm}{\ket{\bar{2}_{-}}}
\newcommand{\dktp}{\ket{\bar{2}_{+}}}
\newcommand{\dku}{\ket{\bar{U}}}
\newcommand{\dkd}{\ket{\bar{D}}}
\newcommand{\dbom}{\bra{\bar{1}_{-}}}
\newcommand{\dbop}{\bra{\bar{1}_{+}}}
\newcommand{\dbtm}{\bra{\bar{2}_{-}}}
\newcommand{\dbtp}{\bra{\bar{2}_{+}}}

\newcommand{\dbd}{\bra{\bar{D}}}

\newcommand{\absp}{\big\lvert lr'+\eta\,l'r\big\rvert^2}
\newcommand{\absm}{\big\lvert lr'-\eta\,l'r\big\rvert^2}

\graphicspath{{figure/}}

\begin{document}
	
	\title{Indistinguishability-enhanced entanglement recovery by spatially localized operations and classical communication}
	
	\author{Matteo Piccolini}
	\email{matteo.piccolini@unipa.it}
	\affiliation{Dipartimento di Ingegneria, Universit\`{a} di Palermo, Viale delle Scienze, 90128 Palermo, Italy}
	\affiliation{INRS-EMT, 1650 Boulevard Lionel-Boulet, Varennes, Qu\'{e}bec J3X 1S2, Canada}
	
	\author{Farzam Nosrati}
	\affiliation{Dipartimento di Ingegneria, Universit\`{a} di Palermo, Viale delle Scienze, 90128 Palermo, Italy}
	\affiliation{INRS-EMT, 1650 Boulevard Lionel-Boulet, Varennes, Qu\'{e}bec J3X 1S2, Canada}
	
	\author{Roberto Morandotti}
	\affiliation{INRS-EMT, 1650 Boulevard Lionel-Boulet, Varennes, Qu\'{e}bec J3X 1S2, Canada}
	
	\author{Rosario Lo Franco}
	\email{rosario.lofranco@unipa.it}
	\affiliation{Dipartimento di Ingegneria, Universit\`{a} di Palermo, Viale delle Scienze, 90128 Palermo, Italy}

\begin{abstract} 

We extend a procedure exploiting spatial indistinguishability of identical particles to recover the spoiled entanglement between two qubits interacting with Markovian noisy environments. Here, the spatially localized operations and classical communication (sLOCC) operational framework is used to activate the entanglement restoration from the indistinguishable constituents. We consider the realistic scenario where noise acts for the whole duration of the process. Three standard types of noises are considered: a phase damping, a depolarizing, and an amplitude damping channel. Within this general scenario, we find the entanglement to be restored in an amount proportional to the degree of spatial indistinguishability. These results elevate sLOCC to a practical framework for accessing and utilizing quantum state protection within a quantum network of spatially indistinguishable subsystems.

\end{abstract}

	\date{\today }

	\maketitle

    \section{Introduction}
    
    Entanglement plays a crucial role towards the development of quantum information, quantum communication and quantum computation technologies\tilde\cite{horodecki,obrienreview,AltmanPRXQuantum,monroe2002quantum,pirandola2015advances}. For this reason, it is fundamental to study and characterize how quantum correlations behaves in different typical real-world scenarios. This requires to take into account the interactions between the system under consideration and the surrounding environment, leading to an open quantum system picture where entanglement is subjected to the detrimental action of ubiquitous noise\tilde\cite{breuer2002theory,suter2016colloquium,preskill_2018,rotter_2015}. Entanglement protection and entanglement recovery techniques are thus crucial in guaranteeing a florid future for quantum technologies. Some of the strategies developed to this aim include quantum error corrections\tilde\cite{preskill_1998,knill2005quantum,shor_1995,steane_1996}, decoherence-free subspaces\tilde\cite{zanardi1997noiseless,lidar1998decoherence}, dynamical decoupling and control techniques\tilde\cite{Viola1998,viola2005random,darrigo_2014_aop,franco2014preserving,orieux_2015,facchi_2004,lo_franco_2012_pra,xu_2013,damodarakurup_2009,cuevas_2017}, distillation protocols\tilde\cite{Bennett1996,kwiat_2001,dong_2008} and structured environments with memory effects\tilde\cite{mazzola_2009,bellomo_2008,lo_franco_2013,aolita_2015,xu_2010,bylicka_2014,man_2015,tan_2010,tong_2010,breuer_2016_colloquium,man_2015_pra}. Nonetheless, each of these approaches comes with its own trade-offs; for example, quantum error correction techniques require to increase the number of qubits and computational steps needed to run the considered process. Dynamical decoupling schemes require additional control operations which can be themselves faulty, so that the process has to be very well-designed to suppress experimental imperfections almost perfectly\tilde\cite{suter2016colloquium}. In general, all the above-mentioned strategies are suited to contrast specific types of errors in specific scenarios while any real-world quantum technology application involves a combination of them.
    
    A recently-developed new strategy exploits a different resource to contrast noise: the indistinguishability of identical particles.
    The importance of investigating identical particles clearly emerges from the fundamental role they play as building blocks of many different physical systems
    employed in quantum information and quantum communication tasks, such as quantum networks, superconducting circuits, quantum dots, Bose-Einstein condensates, optical platforms, and many more.
    Nonetheless, the characterization of quantum correlations between identical constituents is still a debated topic\tilde\cite{ghirardi1977some,ghirardi,horodecki,shi2003quantum,sasaki2011entanglement,morris2020entanglement,benatti2020entanglement}, which has led to a wide pletora of different approaches over time\tilde\cite{ghirardi,facchiIJQI,Li2001PRA,Paskauskas2001PRA,cirac2001PRA,zanardiPRA,eckert2002AnnPhys,balachandranPRL,sasaki2011PRA,giulianoEPJD,bose2002indisting,bose2013,tichyFort,PlenioExtracting,sciaraSchmidt,nolabelappr,compagno2018dealing,slocc,morrisPRX}.
    Among these, the \textit{spatially localized operations and classical communications} (sLOCC) framework has emerged as an operational protocol which exploits spatial indistinguishability to generate entanglement between identical constituents\tilde\cite{slocc,experimentalslocc,lee2021entangling}. In its simplest form, two identical two-level particles with opposite pseudo-spin are initially localized in distinct regions of space, being distinguishable and individually addressable. Later on, a \textit{deformation} process leads the two particles wave functions to spatially overlap over two distinct regions of space. The result is an indistinguishable bipartite state whose degree of spatial indistinguishability can be quantified by a proper entropic measure\tilde\cite{indistentanglprotection}. A particle detection is now carried out on the spatial modes where the wave functions are overlapped; by post-selecting the results where exactly one qubit per location is found, an entangled state is obtained where the constituents are once again distinguishable. Most importantly, the amount of quantum correlations contained in the final state is found to be proportional to the degree of spatial indistinguishability achieved with the deformation, thus showing that indistinguishability of identical constituents provides a key to access entanglement. Remarkably, when the generated spatial overlap is maximum, this procedure produces a maximally entangled state.
    
    \begin{figure*}[t!]
		\includegraphics[width=0.83\textwidth]{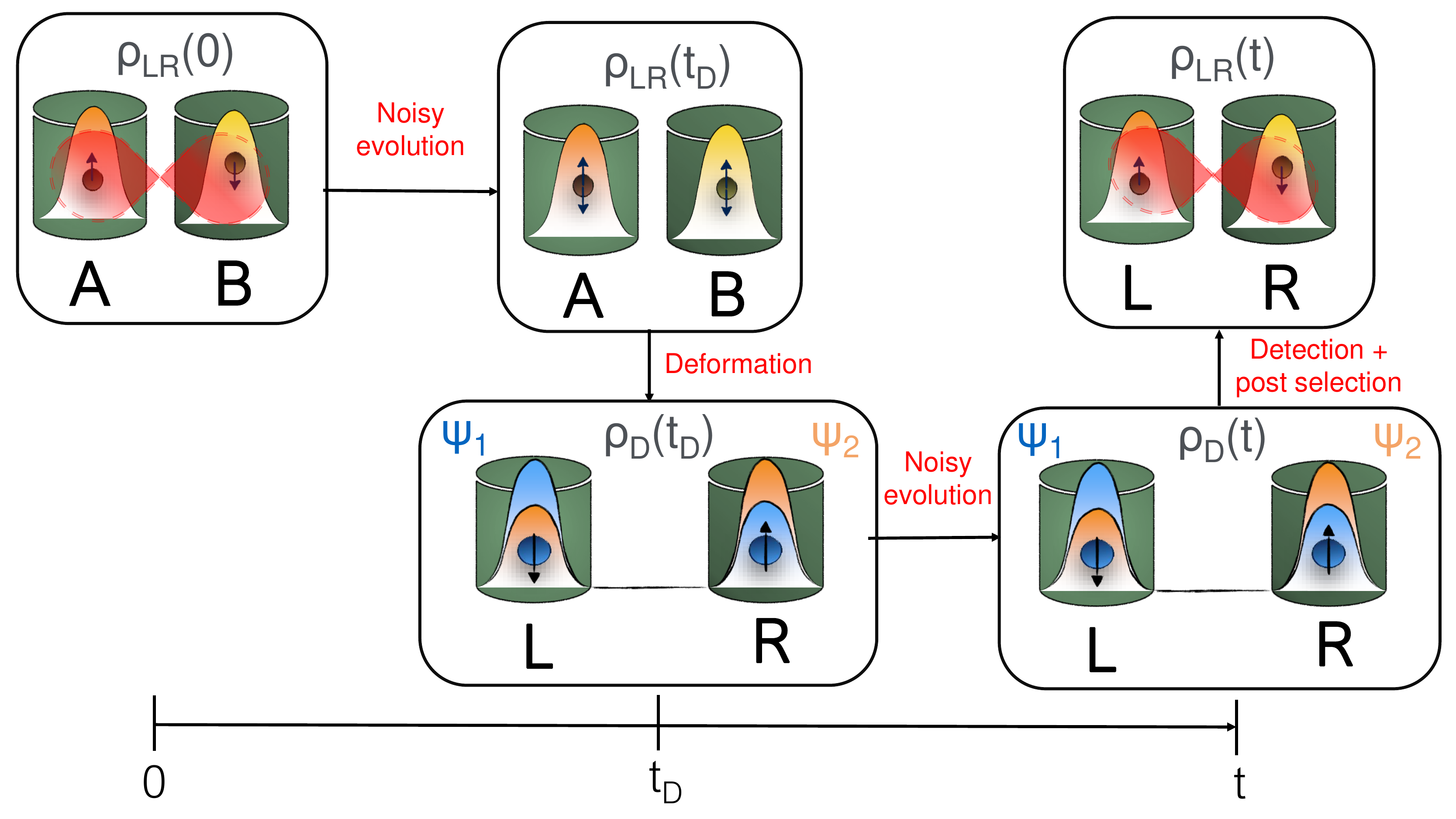}
		\centering
		\caption{\textbf{Illustration of the process.} Two initially distinguishable and entangled qubits are let to interact with two independent noisy environments. At time $\td$ a deformation takes the two particles spatial wave functions to overlap, generating spatial indistinguishability. The noisy interaction with the environments continues until time $t$, when the sLOCC measurement is carried out on the system.}
	\label{process}
	\end{figure*}

    In order to understand how the sLOCC protocol could be exploited to contrast entanglement degradation in an open system scenario, we need to make one further assumption. Indeed, let us suppose that the initial state is maximally entangled. When no noise is involved, the sLOCC protocol described above simply outputs the initial state. Nonetheless, things get more interesting when noise is taken into consideration.
    In Ref.\tilde\cite{indistdynamicalprotection}, the authors analyzed the following scenario: starting with two qubits in a Bell singlet state, the deformation step bringing the two wave functions to spatially overlap is carried out before the particles can interact with their surroundings. After that, two noisy environments of the same type localized on the spatial regions where the indistinguishable constituents have been deformed are let to interact with the two qubits. Three standard types of noisy environments have been considered: a phase damping channel, a depolarizing channel, and an amplitude damping channel. During the interaction, the quantum correlations within the system gets spoiled. Finally, the detection and postselection steps are carried out. Remarkably, the state so obtained shows the same characteristics of the one generated starting from non-entangled constituents: namely, it is an entangled state whose amount of quantum correlations increases with the degree of spatial indistinguishability achieved with the deformation. In particular, when the overlap is maximum the original maximally entangled state is restored. Thus, the sLOCC operational protocol provides an effective procedure not only to \textit{activate} but also to \textit{recover} the quantum correlations initially present within a state subjected to the detrimental action of a standard noisy environment.

    In Ref.\tilde\cite{Piccolini_2021}, the authors provided a similar analysis in a slightly different scenario; in particular, they considered the noisy interaction to occur only before the deformation, assuming the detection to be carried out immediately after the state gets indistinguishable. Even in this case, the sLOCC operational framework has been found to provide an effective entanglement recovery procedure.
    
    In this paper, we aim to join Refs.\tilde\cite{indistdynamicalprotection} and\tilde\cite{Piccolini_2021} so as to consider the more general scenario where noise spoils the initial quantum correlations during the whole sLOCC process, i.e. both before and after the indistinguishability is activated. Our goal is to provide a quantitative analysis of the sLOCC protocol as an effective procedure to restore entanglement in real-world applications.
    In what follows, identical particles are treated within the \textit{no-label} formalism\tilde\cite{nolabelappr,compagno2018dealing,slocc}, a mathematical approach which allows to overcome some of the main problems affecting the standard approach in dealing with entanglement\tilde\cite{tichy,ghirardi}.

    \section{Methods}

	The considered process is displayed in Figure\tilde\ref{process}.
	We begin with two identical qubits localized in two distinct regions of space $A$ and $B$ initially prepared in the pure maximally entangled state $\rab(0)=\kom_\textrm{AB}\bom_\textrm{AB}$, where $\kom_\textrm{AB}=(\kaubd-\kadbu)/\sqrt{2}$ is the usual Bell singlet state. The bipartite system is let to interact with two independent noisy environments localized one on A and one on B.
	At time $\td$, the so obtained (generally) mixed state $\rab(\td)$ is deformed into the indistinguishable state $\rd(\td)$. We recall that the deformation process amounts to modifying the spatial wave functions of two different particles in a tunable way to make them spatially overlapped; this action is described by the deformation operator $\mathcal{D}$ performing the following transformation\tilde\cite{Piccolini_2021}:
	\begin{equation}
	\label{deformation}
	    \ket{A\tau_1}\otimes\ket{B\tau_2}
	    \xrightarrow{\mathcal{D}}
	    \ket{\psi_1\tau_1,\psi_2\tau_2},
	\end{equation}
	where $\psi_1$ and $\psi_2$ are at least partially overlapped spatial modes ($\ket{\psi_1\tau_1,\psi_2\tau_2}\neq\ket{\psi_1\tau_1}\otimes\ket{\psi_2\tau_2}$) while $\tau_j=\uparrow,\downarrow$ is the pseudospin of the $j$-th particle. Notice that, despite the initial spatial modes being orthogonal $\braket{A|B}=0$, the final spatial wave functions in general are not ($\braket{\psi_1|\psi_2}\neq0$), so that the deformation operator in Eq.\tilde(\ref{deformation}) takes normalized states into unnormalized ones and can thus be non-unitary.
	
	For convenience, we follow the same simplification as in Ref.\tilde\cite{indistdynamicalprotection,Piccolini_2021} and assume the deformation to overlap the spatial wave functions over two distinct regions of space $L$ and $R$, so that
	\begin{equation}
	\label{wfstructure}
	    \ket{\psi_1}=l\ket{L}+r\ket{R},\quad
	    \ket{\psi_2}=l'\ket{L}+r'\ket{R}.
	\end{equation}
	Here $\braket{L|R}=0$, while the complex coefficients $l,\,l',\,r,\,r'$ are such that $|l|^2+|r|^2=|l'|^2+|r'|^2=1$.
	
	As a consequence of the achieved state's indistinguishability, the noisy environments are now not able anymore to distinguish the particle they are acting on. The interaction continues until time $t$, when the detection and postselection steps of the sLOCC process are carried out. These amount to projecting the state $\rd(t)$ over the subspace spanned by the basis
	\begin{equation}
	\label{basis}
	    \mathcal{B}_\textrm{LR}=\{\kluru,\klurd,\kldru,\kldrd\},
	\end{equation}
	an operation performed by the projection operator
	\begin{equation}
	\label{sloccprojector}
    	\hat{\Pi}_\textrm{LR}
    	=\sum_{\sigma,\tau=\uparrow,\downarrow}\ket{L\sigma,R\tau}\bra{L\sigma,R\tau}.
	\end{equation}
	The (normalized) result is the final distinguishable state
	\begin{equation}\label{sloccstate}
	    \rlr(t)
	    =\frac{\hat{\Pi}_\textrm{LR}\,\rd(t)\,\hat{\Pi}_\textrm{LR}}{\textrm{Tr}\left[\hat{\Pi}_\textrm{LR}\,\rd(t)\right]},
	\end{equation}
	with one particle localized on $L$ and the other on $R$, postselected with probability
	\begin{equation}
	\label{sloccprob}
	    P_\textrm{LR}(t)
	    =\textrm{Tr}\left[\hat{\Pi}_\textrm{LR}\,\rd(t)\right].
	\end{equation}
	Summing up, the state of Eq.\tilde(\ref{sloccstate}) is the result of a process where noise has continuously acted in a destructive way on the quantum correlations initially present. Our claim is that the amount of entanglement it carries has nonetheless been restored (partially of completely) by the described procedure.
	
	As an entanglement quantifier, we use the Wootters concurrence\tilde\cite{concurrence,indistentanglprotection} for convenience. This is defined as
	\begin{equation}
	\label{concurrence}
	    C(\rlr)=\max \{0,\sqrt{\lambda_4}-\sqrt{\lambda_3}-\sqrt{\lambda_2}-\sqrt{\lambda_1}\}
	\end{equation}
	where $\lambda_i$ are the eigenvalues of $\xi=\rlr\,\rtilde_{\textrm{LR}}$ sorted in decreasing order,  $\rtilde_{\textrm{LR}}=(\sigma^{\textrm{L}}_{y}\otimes\sigma_{y}^{\textrm{R}})\,\rlr^{*}\,(\sigma^{\textrm{L}}_{y}\otimes\sigma_{y}^{\textrm{R}})$, and $\sigma_{y}^{\textrm{L}}$, $\sigma_{y}^{\textrm{R}}$ are the usual Pauli matrix $\sigma_y$ applied, respectively, to the particle in $\textrm{L}$ and in $\textrm{R}$.
	
	The degree of indistinguishability achieved by the deformation operation of Eq.\tilde(\ref{deformation}) can be quantified by an entropic measure defined in terms of the probabilities of finding each particle in each region\tilde\cite{indistentanglprotection}, namely
	\begin{equation}
	\label{indistinguishability}
	    \mathcal{I}=
	    -\dfrac{|l|^2\,|r'|^2}{\mathcal{Z}} \log_2 \dfrac{|l|^2\,|r'|^2}{\mathcal{Z}}
	    -\dfrac{|l'|^2\,|r|^2}{\mathcal{Z}}\log_2 \dfrac{|l'|^2\,|r|^2}{\mathcal{Z}},
	\end{equation}
    where $l=\braket{L|\psi_1},\,r=\braket{R|\psi_1},\,l'=\braket{L|\psi_2},\,r'=\braket{R|\psi_2}$ are the coefficients introduced in Eq.\tilde(\ref{wfstructure}) and $\mathcal{Z}=|l|^2\,|r'|^2+|l'|^2\,|r|^2$. These coefficients are experimentally tunable, so that we set them to be real; in order to reduce the indistinguishability of Eq.\tilde(\ref{indistinguishability}) to a function of one variable only, we impose the further condition $|r|=|l'|$ in what follows.
	
	Finally, we consider the fidelity as an important figure of merit to show that the higher the degree of spatial indistinguishability achieved, the closer the final state of Eq.\tilde(\ref{sloccstate}) will be to the initial Bell singlet one; we recall that for an initial pure state such quantity is simply given by
	\begin{equation}
	\label{fidelity}
	    F\big(\rlr(0),\rlr(t)\big)
	    ={}_\textrm{LR}\bom\rlr(t)\kom_\textrm{LR},
	\end{equation}
	where we have set $A=L,\,B=R$ for the comparison to be meaningful.

	\section{Noisy environments}
	
    In this work, we model the two independent noisy environments as memoryless (Markovian) baths of harmonic oscillators at zero temperature with a single excited mode coupled to the qubit they are interacting with. Within this typical qubit-cavity model, each environment induces a single-particle decay rate $\gamma_0$ and is characterized by the Lorentzian spectral density\tilde\cite{breuer2002theory,Haikka_2010}
	\begin{equation}
	\label{lorentzdensity}
    	J(\omega)=
    	\frac{\gamma_0}{2\pi}
    	\frac{\lambda^2}{(\omega-\omega_0)^2+\lambda^2},
	\end{equation}
	where $\omega_0$ is the qubit transition frequency and $\lambda$ is the spectral width of the reservoir modes.
    The two environments are assumed to be identical, thus being characterized by the same parameters $\gamma_0$ and $\lambda$. Those parameters define two times characteristic of the interaction and of the bath, namely the relaxation time $\tau_R\approx\gamma_0^{-1}$ over which the state of the system changes significantly and the bath correlation time $\tau_B\approx\lambda^{-1}$. Our request of Markovianity translates in the condition $\tau_R\gtrsim2\tau_B$.
    
    Three standard noisy environments are considered: a phase damping channel, a depolarizing channel and an amplitude damping channel.
    Given the nature of the process, the analysis can be broken down into two parts with respect to the generation of the spatial indistinguishability. 
    
        \subsection{Distinguishable particles: Kraus operators approach}
        
        When the qubits are still distinguishable, we determine their dynamics via the typical Kraus operators formalism\tilde\cite{nielsen2010quantum}.
        The system density matrix $\rho_s$ evolves according to
        \begin{equation}
        \label{kraus}
            \rho_s(t)=\sum_{i,j} E_{ij} \rho_s(0) E_{ij}^\dagger,
        \end{equation}
        where $E_{ij}:=E_i^L\otimes E_j^R$ while $E_i^X$ are the time-dependent Kraus operators specific of the channel under consideration, acting on the qubit in the region $X=L,R$. Such operators encode the time-dependent disturbance probability $p(t)$ introduced by the the environments on the system. This is actually a decoherence function which can be computed by solving the differential equation\tilde\cite{breuer2002theory,Bellomo_2007}
    	\begin{equation}
    	\label{probability}
    	    \dot{q}(t)
    	    =-\int_{0}^{t}dt_1\,f(t-t_1)\,q(t_1),
    	\end{equation}
        where $p(t)=1-q(t)$.
        Here, the correlation function $f(t-t_1)$ is determined by the Fourier transform of the spectral density of the bath
        \begin{equation}
    	    f(t-t_1)=
    	    \int d\omega\,J(\omega)\,e^{-i(\omega-\omega_0)(t-t_1)}.
    	\end{equation}
    	The disturbance probability induced by an environment featuring the Lorentzian spectral density of Eq.\tilde(\ref{lorentzdensity}) is easily computed to be\tilde\cite{breuer2002theory}
    	\begin{equation}
    	\label{probabilityfunc}
    	    p(t)
    	    =1-e^{-\lambda t}\left[\cos\left(\frac{d\,t}{2}\right)+\frac{\lambda}{d}\sin\left(\frac{d\,t}{2}\right)\right]^2,
    	\end{equation}
    	with $d:=\sqrt{2\gamma_0\lambda-\lambda^2}$.
	
    	\subsection{Indistinguishable particles: master equation approach}
    	
    	After the deformation, the Kraus operators approach becomes harsh due to the practical difficulty in finding an explicit expression for the $E_{ij}$ coefficients of Eq.\tilde(\ref{kraus}) when the particles are indistinguishable to the environment\tilde\cite{indistdynamicalprotection}; indeed, now the relation $E_{ij}\neq E_i^L\otimes E_j^R$ is not valid anymore\tilde\cite{indistdynamicalprotection}. For this reason, we resort to a master equation approach\tilde\cite{breuer2002theory,lindblad1976generators,gorini1976completely}.
    	
    	Before proceeding with the scenario of interest, let us consider the simpler situation of one single qubit $\rho_S$ interacting with an environment $\rho_B$. Let us suppose that the following assumptions hold:
    	\begin{enumerate}
    	    \item 
    	    (\textit{Born approximation})
    	    the coupling is weak, so that the environment is negligibly affected by the interaction and the global state $\rho_{\textrm{tot}}$ at time $t$ can be approximated by the product state $\rho_{\textrm{tot}}(t)\approx\rho_S(t)\otimes\rho_B$;
    	    
    	    \item
    	    (\textit{Markovian approximation})
    	    the regime is Markovian, so that the evolution of the system state can be made local in time;
    	    
    	    \item
    	    the total Hamiltonian in the Schrodinger picture is given by $H_T=H_S+H_B+H_I$, where the system free Hamiltonian $H_S$ and the environment free Hamiltonian $H_B$ are time independent while the Hamiltonian $H_I$ encodes the interaction between the qubit and the environment;
    	    
    	    \item
    	    the environment is a bath of harmonic oscillators whose Hamiltonian is given by $H_B=\sum_k \omega_k a_k^\dagger a_k$, with $a_k$ and $a_k^\dagger$ being the annihilation and creation operators of the mode $k$ having frequency $\omega_k$;
    	    
    	    \item
    	    the interaction Hamiltonian has the form 
    	    \begin{equation}
    	    \label{interactionhamiltgeneral}
    	        H_I=S\otimes B^\dagger+S^\dagger\otimes B,
    	    \end{equation}
        	where $S$ and $B$ are operators acting only on the system and on the environment, respectively;
    	    
    	    \item
    	    the operator B in Eq.\tilde(\ref{interactionhamiltgeneral}) has the form $B=\sum_k g_k^* a_k$;
    	    
    	    \item
    	    the bath is at zero temperature and $\rho_B$ is the vacuum state;
            
            \item
            the bath is characterized by the Lorentzian spectral density of Eq.\tilde(\ref{lorentzdensity});
            
            \item
            there is just one resonant frequency.
    	\end{enumerate}
        Under these assumptions, the dynamics of the system is given by the master equation (in the interaction picture)\tilde\cite{brasil2013simple,manzano2020short}
        \begin{equation}
        \label{mastereq}
            \frac{d}{dt}\rho_S(t)
            =\gamma_0\left[
            S\,\rho_S(t)\,S^\dagger-\frac{1}{2}\big\{S^\dagger S,\,\rho_S (t)\}
            \right].
        \end{equation}
        All the conditions expressed above hold in our scenario; in particular:
        \begin{itemize}
            \item
            conditions $1$ and $2$ \textit{define} the regime we are working in, as stated at the beginning of this section. We recall that, when accompanied by the \textit{rotating wave approximation}, these assumption underlie the well-known Gorini–Kossakowski–Sudarshan–Lindblad (GKLS) quantum optical master equation\tilde\cite{breuer2002theory};
            
            \item
            conditions $3,\,4,\,6,\,7,\,8,$ and $9$ are characteristics of our model;
            
            \item
            condition $5$ holds for the noisy channels we are considering, as we will show later.
        \end{itemize}
        Going back to our scenario, we conclude that Eq.\tilde(\ref{mastereq}) would dictate the evolution of both qubits \textit{if} they were distinguishable to the eyes of the environments. Nonetheless, the spatial indistinguishability generated by the deformation implies that we need to find a suitable generalization of the master equation for the dynamics of $\rd$. A generalization of the interaction Hamiltonian between the indistinguishable bipartite system and the environments is given by\tilde\cite{indistdynamicalprotection}
        \begin{equation}
        \label{interactionhamilt}
            H_I
            =S_L\otimes\sum_k g_{kL}\,a_{kL}^\dagger
            +S_R\otimes\sum_k g_{kR}\,a_{kR}^\dagger+h.c.,
        \end{equation}
        where $S_{X}$ is a generic spatially localized single particle operator acting on the pseudospin of the qubit in the region $X=L,R$ only, $g_{kX}$ are the coupling constants between the qubit and the mode $k$ of the environment in region $X$, and \textit{h.c.} indicates the Hermitian conjugation.
        The Hamiltonian of Eq.\tilde(\ref{interactionhamilt}) clearly shows a structure of the form $H_I=H_L+H_R$, where $H_X$ acts only on the region $X=L,R$. Since the two environments are distinct, the operators $a_{kX}$ only affect the bath localized in the corresponding region $X$. On the contrary, the two particles are spatially distributed on both $L$ and $R$, so that the two pseudospin operators $S_L$ and $S_R$ affect both qubits. To be physically meaningful, such action must be weighted by the probability amplitude associated to each particle of being in the region $X$. This is encoded within the definition of the action of a single particle operator $O_X$ spatially localized on $X$ acting on a state of N identical particles $\ket{\Psi}=\ket{\psi_1,\,\psi_2,\dots,\psi_N}$ within the no label approach, which is given by\tilde\cite{indistdynamicalprotection}
        \begin{equation}
        \label{singleopaction}
            O_X\ket{\Psi}
            :=\sum_i\,\lvert\braket{X|\psi_i}\rvert\ket{\psi_1,\dots,O\,\psi_i,\dots,\psi_N}.
        \end{equation}
        Using this definition, it immediately follows by acting on a test state that the interaction Hamiltonian of Eq.\tilde(\ref{interactionhamilt}) can be rewritten as
        \begin{equation}
        \label{interactionhamilt2}
            H_I=\sum_{X=L,R}\,\sum_{j=1,2} H_{I,X}^{(j)},
        \end{equation}
        where
        \begin{equation}
        \label{interactionhamilt3}
            H_{I,X}^{(j)}
            =S_X^{(j)}\otimes\sum_k\Big(\lvert\braket{X|\psi_j}\rvert\,g_{kX}\Big)\,a_{kX}^\dagger
            \,+\,h.c.
        \end{equation}
        and the superscript $(j)$ indicates that the corresponding operator acts only on the single particle state at the j-th slot of the global system state vector. It is now easy to notice that each term of Eq.\tilde(\ref{interactionhamilt3}) in the interaction Hamiltonian of Eq.\tilde(\ref{interactionhamilt2}) acts on a single particle and is of the form of Eq.\tilde(\ref{interactionhamiltgeneral}), with the system operators $S$ given by the single particle operators $S_X^{(j)}$ and the coupling constants $g_{kX}$ replaced by the \textit{effective} coupling constants $\lvert\braket{X|\psi_j}\rvert\,g_{kX}$. By following the standard procedure that leads to Eq.\tilde(\ref{mastereq}) with these small modifications, we get the desired master equation for the reduced density matrix of two indistinguishable qubits in the interaction picture\tilde\cite{indistdynamicalprotection}:
            \begin{equation}
            \label{indistmastereq}
                \frac{d\rho_S(t)}{dt}
                =\sum_{X=L,R}\,\sum_{i,j=1,2}
                \gamma_X^{(i,j)}
                \left[
                    S_X^{(i)}\,\rho_S(t)\,S_X^{(j)\dagger}-\frac{1}{2}\big\{S_X^{(i)^\dagger} S_X^{(j)},\,\rho_S (t)\big\}
                \right],
            \end{equation}
        where $\gamma_X^{(i,j)}:=\gamma_0\,\lvert\braket{X|\psi_i}\braket{\psi_j|X}\rvert$ $(j=1,2)$ are \textit{effective} decay rates. It is important to highlight a main difference between Eq.\tilde(\ref{indistmastereq}) and Eq.\tilde(\ref{mastereq}) which is a consequence of the spatial indistinguishability: the dynamical contribution of the terms with $i\neq j$. It is immediate to notice that when there is no indistinguishability, e.g. when $\ket{\psi_i}=\ket{L}$ and $\ket{\psi_j}=\ket{R}$ ($i\neq j$), the only nonzero effective decay rates are $\gamma^{(1,1)}=\gamma^{(2,2)}=\gamma_0$ and Eq.\tilde(\ref{indistmastereq}) reduces to the master equation of two distinguishable particles.

    \section{Results}
    \label{results}
    
    In this section we report the results obtained for the phase damping channel, the depolarizing channel, and the amplitude damping channel. Before moving on, it is useful to introduce the notation used to assess the states involved. Using the no label approach, we define:
    \begin{itemize}
        \item
        distinguishable particles states:
        \begin{equation}
        \label{diststates}
            \begin{gathered}
                \ket{1_\pm}_{\textrm{AB}}
                :=\frac{1}{\sqrt{2}}\Big(\kaubd\pm\kadbu\Big),\\
                \ket{2_\pm}_{\textrm{AB}}
    	        :=\frac{1}{\sqrt{2}}\Big(\kaubu\pm\kadbd\Big),\\
    	        \ket{U}_\textrm{AB}
    	        :=\kaubu,
    	        \qquad
    	        \ket{D}_\textrm{AB}
    	        :=\kadbd.
            \end{gathered}
        \end{equation}
        We gather the states of Eq.\tilde(\ref{diststates}) into two orthonormal basis of the distinguishable Hilbert space, namely
        \begin{equation}
        \label{distbasis}
            \begin{gathered}
                \mathcal{B}_1
                :=\Big\{
                    \kop_\textrm{AB},\,\kom_\textrm{AB},\,\ktp_\textrm{AB},\,\ktm_\textrm{AB}
                \Big\},
                \\
                \mathcal{B}_2
                :=\Big\{
                    \kop_\textrm{AB},\,\kom_\textrm{AB},\,\ket{U}_\textrm{AB},\,\ket{D}_\textrm{AB}
                \Big\}
                ;
            \end{gathered}
        \end{equation}
        
        \item
        indistinguishable particles states:
        \begin{equation}
        \label{indiststates}
            \begin{gathered}
                \ket{\bar{1}_\pm}
                :=\frac{1}{\sqrt{2}}\Big(\ket{\psi_1\uparrow,\psi_2\downarrow}\pm\ket{\psi_1\downarrow,\psi_2\uparrow}\Big),\\
                \ket{\bar{2}_\pm}
                :=\frac{1}{\sqrt{2}}\Big(\ket{\psi_1\uparrow,\psi_2\uparrow}\pm\ket{\psi_1\downarrow,\psi_2\downarrow}\Big),\\
                \dku:=\ket{\psi_1\uparrow,\psi_2\uparrow},
                \qquad
                \dkd:=\ket{\psi_1\downarrow,\psi_2\downarrow},
            \end{gathered}
        \end{equation}
        where $\psi_1$ and $\psi_2$ are the single particle distributed wave functions defined in Eq.\tilde(\ref{wfstructure}).
        States of Eq.\tilde(\ref{indiststates}) are the result of the action of the deformation given by Eq.\tilde(\ref{deformation}) on the states of Eq.\tilde(\ref{diststates}). As previously mentioned, these are in general unnormalized (but orthogonal) states; in particular, computing their inner product we get
            \begin{align}
            \label{statesnorm}
                    \braket{\bar{1}_{+}|\bar{1}_{+}}
                    &=\braket{\bar{2}_{+}|\bar{2}_{+}}
                    =\braket{\bar{2}_{-}|\bar{2}_{-}}\nonumber\\
                    &=\braket{\bar{U}|\bar{U}}
                    =\braket{\bar{D}|\bar{D}}=C_{+}^2,\nonumber
                    \\
                    \braket{\bar{1}_{-}|\bar{1}_{-}}&=C_{-}^2,
            \end{align}
        where
        \begin{equation}
        \label{coefficients}
            C_\pm:=\sqrt{1\pm\eta\lvert\braket{\psi_1|\psi_2}\rvert^2}.
        \end{equation}
        Normalized, indistinguishable states can be simply obtained by means of these coefficients. Nonetheless, when computing the concurrence we are only interested in the distinguishable final states, so that we can disregard the normalization of the indistinguishable states and limit ourselves to perform the normalization after the sLOCC projection (see Eq.\tilde(\ref{sloccstate})).
        We gather the states of Eq.\tilde(\ref{indiststates}) into two orthogonal basis of the indistinguishable Hilbert space as follows:
        \begin{equation}
        \label{indistbasis}
            \begin{gathered}
                \mathcal{\bar{B}}_1
                :=\Big\{
                    \dkop,\,\dkom,\,\dktp,\,\dktm,
                \Big\},
                \\
                \mathcal{\bar{B}}_2
                :=\Big\{
                    \dkop,\,\dkom,\,\dku,\dkd
                \Big\}.
            \end{gathered}
        \end{equation}
        \end{itemize}

        \subsection{Phase damping channel}
        The phase damping channel models a nondissipative noise leading to the degradation of quantum coherence without loss of energy. A system interacting with an environment of this type thus preserves its energy eigenstates while losing the interference terms. Physical systems where such a phenomena occurs are, e.g., random telegraph noise and phase noisy lasers \cite{zhou,lo_franco_2012,bordone,bellomo_2012,cai,wold}, photons randomly scattering through a waveguide, and superconducting qubits under low-frequency noise.
        
        The single particle Kraus operators characterizing the phase damping channel are\tilde\cite{breuer2002theory}
        \begin{equation}
	    \label{krauspdc}
	        \begin{gathered}
            	E_0
            	=            	\sqrt{1-p(t)}\,\ket{\uparrow}\bra{\uparrow}
            	+	\ket{\downarrow}\bra{\downarrow}
            	=E_0^\dagger
            	,\\
            	E_1
            	=\sqrt{p(t)}\,\ket{\uparrow}\bra{\uparrow}=E_1^\dagger,
	        \end{gathered}
	    \end{equation}
	    where $p(t)$ is given by Eq.\tilde(\ref{probabilityfunc}).
        Evolving the initial state $\rab(0)=\kom_\textrm{AB}\bom_\textrm{AB}$ with these operators until time $\td$ according to Eq.\tilde(\ref{kraus}), we get
            \begin{equation}
        	\label{distpd}
            	\rab(\td)
            	=\frac{p(\td)}{2}\kop_\textrm{AB}\bop_\textrm{AB}
            	+\left(1-\frac{1}{2}\,p(\td)\right)\kom_\textrm{AB}\bom_\textrm{AB}.
        	\end{equation}
    	
    	At time $\td$ we perform the deformation of Eq.\tilde(\ref{deformation}) which simply maps $\kop_\textrm{AB}\xrightarrow{\mathcal{D}}\dkop$ and $\kom_\textrm{AB}\xrightarrow{\mathcal{D}}\dkom$, obtaining the state
    	\begin{equation}
	    \label{indistpd}
        	\rho_\textrm{D}(\td)
        	=\frac{p(\td)}{2}\dkop\dbop
        	+\left(1-\frac{1}{2}\,p(\td)\right)\dkom\dbom.
	    \end{equation}
	    The interaction Hamiltonian for the indistinguishable qubits interacting with the phase damping channels is given by Eq.\tilde(\ref{interactionhamilt2}) with
	    \begin{equation}
	    \label{interactionpd}
	        H_{I,X}^{(j)}= S_z^{(j)}\otimes\sum_k \Big(\lvert\braket{X|\psi_j}\rvert\,g_{kX}\Big) a_{kX}^\dagger + h.c.,
	    \end{equation}
	    where $S_z=\frac{1}{2}\sigma_z$ is the projection of the spin operator along the $z$ axis given in terms of the usual Pauli matrix $\sigma_z$.
	    By comparing Eq.\tilde(\ref{interactionpd}) with Eq.\tilde(\ref{interactionhamilt3}) we immediately see that the system operator is $S_X=S_X^\dagger=S_z$, so that the system's dynamics is given by the master equation of Eq.\tilde(\ref{indistmastereq})
    	    \begin{equation}
            \label{pdmastereq}
                \frac{d}{dt}\rho_\textrm{D}(t)
                =\sum_{X=L,R}\sum_{i,j=1,2}
                \gamma_X^{(i,j)}
                \left[
                    S_z^{(i)}\,\rho_S(t)\,S_z^{(j)}-\frac{1}{2}\big\{S_z^{(i)} S_z^{(j)},\,\rho_S (t)\big\}
                \right].
            \end{equation}
        Such an evolution preserves the diagonal structure of a state written on the Bell-state basis $\mathcal{\bar{B}}_1$ defined in Eq.\tilde(\ref{indistbasis}), so that we will evolve the state of Eq.\tilde(\ref{indistpd}) on this basis.
        Writing
        \begin{equation}
        \label{pdindistfinalstate}
            \rd(t)=\sum_{u\in\mathcal{\bar{B}}_1}p_u(t)\,\ket{u}\bra{u}
        \end{equation}
        with $\sum_{u\in\mathcal{\bar{B}}_1}p_u(t)=1$, Eq.\tilde(\ref{pdmastereq}) reduces to the system of ordinary differential equations (ODEs)
        \begin{gather}
        \label{pdode}
            \left\{
            \begin{gathered}
                \dot{p}_{1_+}(t)
                =\gamma_-\Big(p_{1_-}(t)-p_{1_+}(t)
                \Big)
                \\
                \dot{p}_{1_-}(t)
                =\gamma_-\Big(p_{1_+}(t)-p_{1_-}(t)
                \Big)
                \\
                \dot{p}_{2_+}(t)
                =\gamma_+\Big(p_{2_-}(t)-p_{2_+}(t)
                \Big)
                \\
                \dot{p}_{2_-}(t)
                =\gamma_+\Big(p_{2_+}(t)-p_{2_-}(t)
                \Big),
            \end{gathered}
            \right.
        \end{gather}
        where we introduced the decay rates
        \begin{equation}
            \gamma_\pm
            :=\sum_{X=L,R}\,\sum_{i,j=1,2} (\pm 1)^{i+j}\,\gamma_0\,\lvert\braket{X|\psi_i}\braket{\psi_j|X}\rvert.
        \end{equation}
	    By solving the above system we get the following equations for the evolution of the populations:
    	\begin{widetext}    
    	    \begin{equation}
    	    \label{pdsolution}
    	        \left\{
    	        \begin{gathered}
        	        p_{1_\pm}(t)
        	        =\frac{1}{2}\bigg[\Big(1+e^{-\gamma_-(t-\td)/2}\Big)p_{1_\pm}(\td)+\Big(1-e^{-\gamma_-(t-\td)/2}\Big)p_{1_\mp}(\td)\bigg]
        	        \\
        	        p_{2_\pm}(t)
        	        =\frac{1}{2}\bigg[\Big(1+e^{-\gamma_+(t-\td)/2}\Big)p_{2_\pm}(\td)+\Big(1-e^{-\gamma_+(t-\td)/2}\Big)p_{2_\mp}(\td)\bigg].
    	        \end{gathered}
    	        \right.
    	    \end{equation}
    	\end{widetext}
	    Looking at Eq.\tilde(\ref{indistpd}), we set $p_{1_+}(\td)=\frac{1}{2}\,p(\td),\,p_{1_-}(\td)=1-\frac{1}{2}\,p(\td),\,p_{2_+}(\td)=0,\,p_{2_-}=0$, and evolve the state according to Eq.\tilde(\ref{pdsolution}). We finally perform the sLOCC measurement of Eq.\tilde(\ref{sloccstate}), obtaining the final normalized distinguishable state
	    \begin{equation}
	    \label{nondissipativefinalstate}
	        \begin{aligned}
	            \rlr(t)
	            =\frac{1}{N}\bigg[&\absp
	            \sum_{u=1_+,2_\pm}p_u(t)\ket{u}_\textrm{LR}\bra{u}_\textrm{LR}\\
	            +&\absm p_{1_-}(t)\kom_{\textrm{LR}}\bom_{\textrm{LR}}\bigg],
	        \end{aligned}
	    \end{equation}
	    with $p_{1_\pm}(t),\,p_{2_\pm}(t)$ given by Eq.\tilde(\ref{pdsolution}) and
	    \begin{equation}
	    \label{nondissipativeN}
	         N:=\absp\sum_{u=1_+,2_\pm}p_u(t)+\absm p_{1_-}(t).
	    \end{equation}

	    \begin{figure}[t]
		    \centering
		    \includegraphics[width=0.49\textwidth]{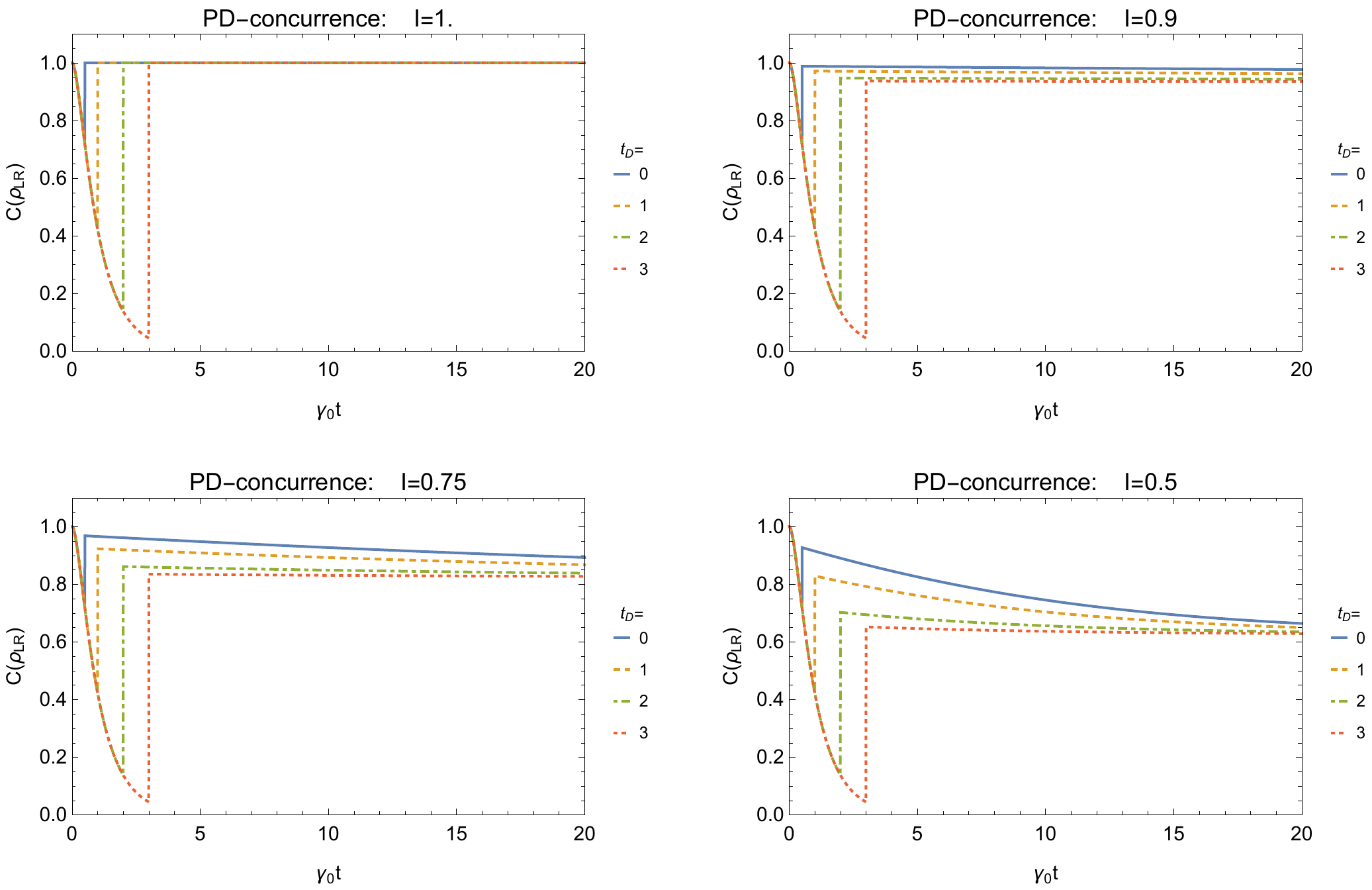}
		    \caption{Concurrence $C(\rho_\mathrm{LR})$ for the case of separated phase damping channels as a function of the total dimensionless interaction time $\gamma_0 t$.
		    Four different deformation times $\gamma_0 t_D$ for four values of spatial indistinguishability $\mathcal{I}$ are reported, under the constraint $|l'|=|r|$. Results hold for fermions with $l,r,l',$ and $r'$ positive and for bosons with one of such coefficients negative.}
		\label{pdconc}
	    \end{figure}

Figure\tilde\ref{pdconc} reports the evolution of the concurrence of Eq.\tilde(\ref{concurrence}) for the two qubits system.
In particular, for $t<\td$ the plots show the concurrence of the distinguishable state of Eq.\tilde(\ref{distpd}) \textit{before} the deformation has occured, while for $t\geq\td$ the concurrence of the state of Eq.\tilde(\ref{nondissipativefinalstate}) is depicted. The shown results hold for fermions with $l,\,r,\,l',\,r'>0$ and for bosons with one of such coefficients negative. Four different degrees of spatial indistinguishability $\mathcal{I}$ are reported, in order to highlight its fundamental role in the entanglement recovery process. Indeed, quantum correlations are clearly shown to rapidly decay with time due to the detrimental action of the environment, before being restored by the deformation at time $\td$ in an amount which increases with $\mathcal{I}$. Furthermore, the different curves for each plot show that such an amount depends also on $\td$ and $t-\td$: the longer we wait to perform the deformation and the sLOCC measurement after it, the lower is the entanglement recovered. Nonetheless, such lowering has a limit: the presence of an horizontal asymptote at a value increasing with the degree of spatial indistinguishability indicates that our procedure allows for an entanglement recovery no matter how long the environmental noise affects the system both before the deformation and between the deformation and the sLOCC detection, \textit{provided that the spatial wave functions have been overlapped at least partially}. Furthermore, when such overlap is maximum (i.e. $\mathcal{I}=1$) our procedure allows for a complete entanglement recovery no matter the interaction time, resuming the total state protection from noise reported in\tilde\cite{indistdynamicalprotection} and\tilde\cite{Piccolini_2021}.
	    
	    	    \begin{figure}[t!]
		    \centering
		    \includegraphics[width=0.49\textwidth]{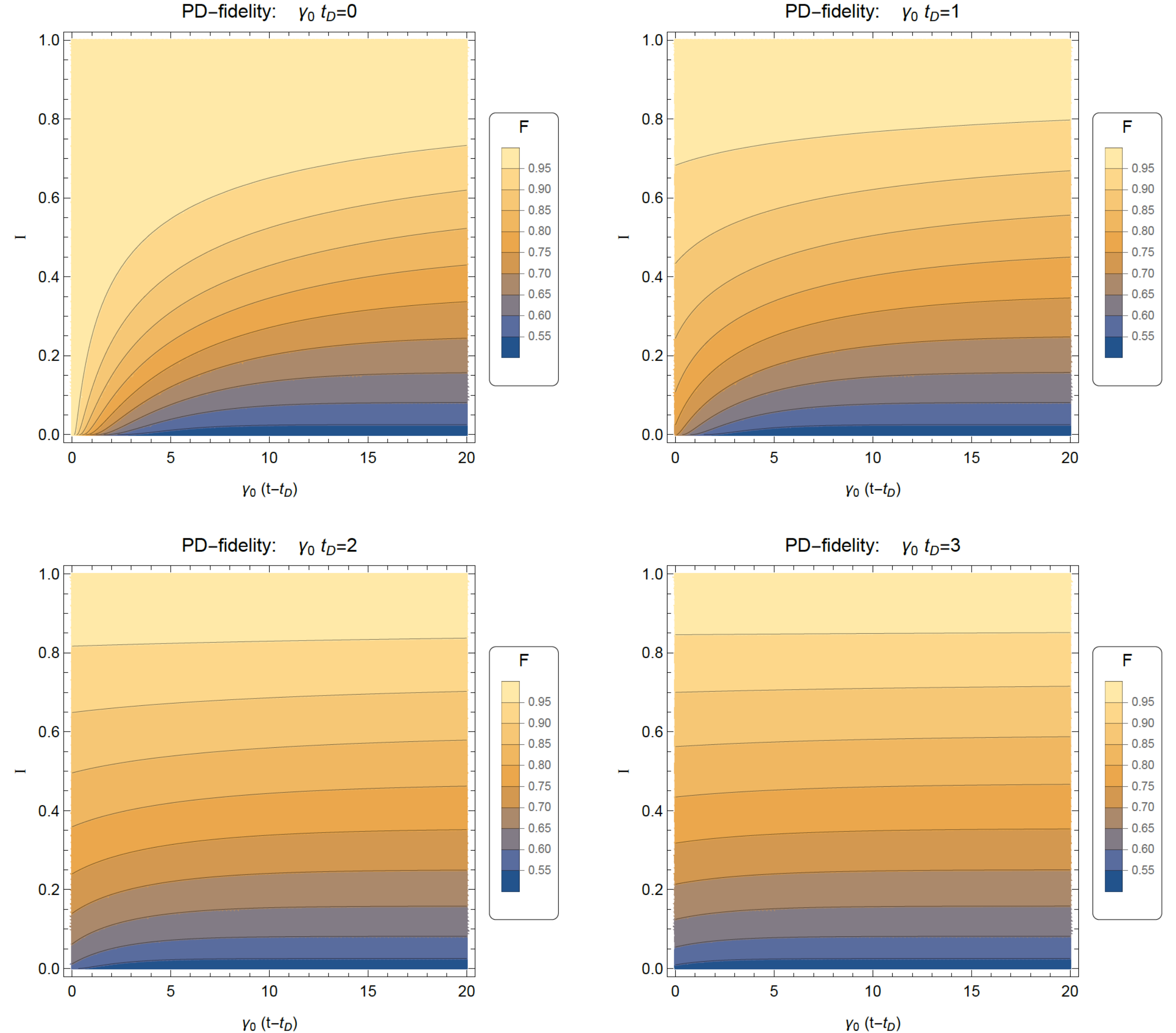}
		    \caption{\footnotesize Fidelity $F$ between final state and initial Bell singlet state in the phase damping scenario versus dimensionless interaction time after the deformation $\gamma_0 (t-t_D)$ and $\mathcal{I}$.
		    Four different interaction times are considered. Results hold for fermions with $l,r,l',$ and $r'$ positive and for bosons with one of such coefficients negative.}
		\label{pdfid}
	    \end{figure}
	    
	    In Figure\tilde\ref{pdfid} we depict the fidelity of Eq.\tilde(\ref{fidelity}) as a function of the dimensionless interaction time after the deformation and the degree of spatial indistinguishability for four different deformation times. As can be noticed, the behaviour of the concurrence is reflected in the fidelity being higher for larger degrees of spatial indistinguishability at fixed times, and decreasing with time at fixed $\mathcal{I}$. The complete protection from noise when $\mathcal{I}=1$ here manifests itself in the fidelity being equal to 1 independently on the interaction time, meaning that the sLOCC protocol has recovered the initial state $\kom$.

	    Finally, we are interested in the probability of success of the sLOCC postselection, given by Eq.\tilde(\ref{sloccprob}). Notice that, in order for Eq.\tilde(\ref{sloccprob}) to actually represent a probability, it is fundamental that the indistinguishable state $\rd(t)$ is normalized, while the state of Eq.\tilde(\ref{pdindistfinalstate}) is not. Thus, we compute the probability of success of the sLOCC postselection as $P_\textrm{LR}(t)
	    =\textrm{Tr}\left[\hat{\Pi}_\textrm{LR}\,\rd(t)\right]/\textrm{Tr}\left[\rd(t)\right]$. Using Eq.\tilde(\ref{statesnorm}), this is easily found to be
	    \begin{equation}
	       \label{nondissipativeprob}
	        P_\textrm{LR}(t)
	        =\frac{
	        N}
	        {
	        C_+^2\,\Big(p_{1_+}(t)+p_{2_+}(t)+p_{2_-}(t)\Big)+C_-^2\,p_{1_-}(t)},
	    \end{equation}
	    where the populations are given in Eq.\tilde(\ref{pdsolution}) and $N$ in Eq.\tilde(\ref{nondissipativeN}).
	    
	    \begin{figure}[t!]
		    \centering
		    \includegraphics[width=0.49\textwidth]{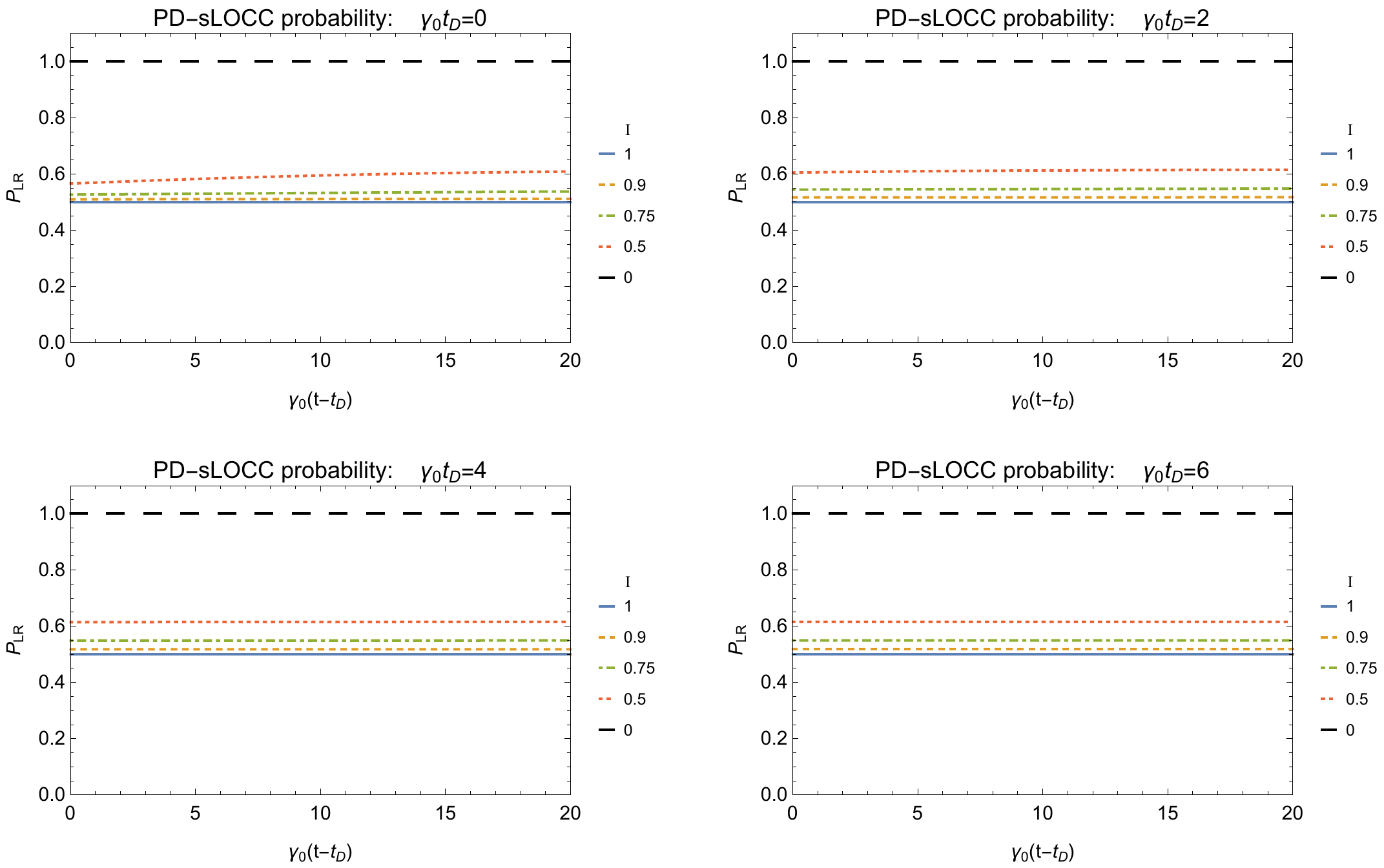}
		    \caption{\footnotesize Postselection (sLOCC) probability $P_\mathrm{LR}$ for the case of separated phase damping channels as a function of the dimensionless interaction time after the deformation $\gamma_0 (t-t_D)$. Five different values of $\mathcal{I}$ for four different deformation times $\gamma_0 t_D$ are reported, under the constraint $|l'|=|r|$. Results hold for fermions with $l,r,l',$ and $r'$ positive and for bosons with one of such coefficients negative.}
		\label{pdprob}
	    \end{figure}
	    
	    Figure\tilde\ref{pdprob} shows the probability of Eq.\tilde(\ref{nondissipativeprob}) as a function of the dimensionless interaction time after the deformation for different values of spatial indistinguishability and deformation times. As can be noticed, there is a trade-off between the concurrence and the probability: indeed, the higher is the first, the lower is the second, meaning that a greater amount of correlations restored comes at the price of discarding more states during the postselection. In particular, the probability of success reaches its minimum value $P_\textrm{LR}=0.5$ when $\mathcal{I}=1$ and its maximum $P_\textrm{LR}=1$ when $\mathcal{I}=0$, i.e. when no deformation is performed and the two qubits remain localized in distinct regions. Finally, the middle-range behaviour $0<\mathcal{I}<1$ sees $P_\textrm{LR}(t)$ reaching asymptotically a value between $0.5$ and $1$ as the interaction time goes to infinity.

	    \subsection{Depolarizing channel}
	    The depolarizing channel models the nondissipative symmetric decoherence which can occur, e.g., in Bose-Einstein condensates\tilde\cite{kasprzak2006bose,zipkes2010trapped}, in the scattering of randomly polarized photons\tilde\cite{puentes2005experimental,puentes2007entangled,shaham2011realizing}, or in nuclear magnetic resonance setups\tilde\cite{xin2017quantum,ryan2009randomized}. A first analysis of the exploitation of spatial indistinguishability as a tool to recover entanglement between two qubits subjected to a depolarizing channel has been carried out in\tilde\cite{indistentanglprotection}, where noise was considered as acting only after the deformation.
	    
	    The action of a depolarizing channel on two distinguishable qubits in the state $\kom_\textrm{AB}$ is known to leave the system untouched with a probability $1-p(t)$ while introducing a white noise with probability $p(t)$ \cite{nielsen2010quantum}. This produces the well-known Werner state which, immediately before the deformation at time $\td$, is given by
	    \begin{equation}
	    \label{wernerdist}
    	    \begin{gathered}
        	    \rab(\td)
        	    =W^-_\textrm{AB}(\td)\\
        	    :=\Big[1-p(\td)\Big]\kom_\textrm{AB}\bom_\textrm{AB}
        	    +\frac{1}{4}\,p(\td)\,\id,
        	\end{gathered}
	    \end{equation}
	    where $1\!\!1=\sum_{j=1_\pm,2_\pm}\ket{j}_\textrm{AB}\bra{j}_\textrm{AB}$ is the $4\times4$ identity operator.
	    
	    We now apply the deformation of Eq.\tilde(\ref{deformation}) to the state of Eq.\tilde(\ref{wernerdist}), obtaining
	    \begin{equation}
	    \label{wernerindist}
	        \begin{aligned}
            	\bar{W}_\textrm{D}^-(\td)
            	:&=\bigg[1-\frac{3}{4}\,p(\td)\bigg]\,\dkom\dbom\\
            	&+\frac{1}{4}\,p(\td)\,\Big[\dkop\dbop+\dktp\dbtp+\dktm\dbtm\Big].
            \end{aligned}
	    \end{equation}
	    The interaction Hamiltonian is given by Eq.\tilde(\ref{interactionhamilt2}) with
	    \begin{equation}
	    \label{interactiondep}
    	    H_{I,X}^{(j)}
    	    =\Big(S_x^{(j)}+S_y^{(j)}+S_z^{(j)}\Big)
    	    \otimes
    	    \sum_k\Big(\lvert\braket{X|\psi_j}\rvert\,g_{kX}\Big)\,a_{kX}^\dagger+h.c.,
	    \end{equation}
	    showing that the system operator $S_X=S_X^\dagger$ is given by the sum of the components $S_{x,y,z}=\frac{1}{2}\sigma_{x,y,z}$ of the spin operator, with $\sigma_{x,y,z}$ being the usual Pauli matrices. The dynamics is dictated by the master equation of Eq.\tilde(\ref{indistmastereq}), which reduces to the system of ODEs
	    \begin{widetext}
    	    \begin{equation}
            \label{depode}
                \left\{
                \begin{aligned}
                    \dot{p}_{1_+}(t)
                    &=\frac{1}{4}\Big[
                    -(2\gamma_++\gamma_-)\,p_{1_+}(t)
                     +\gamma_-\,p_{1_-}(t)
                     +\gamma_+\,p_{2_+}(t)
                     +\gamma_+\,p_{2_-}(t)
                    \Big]
                    \\
                    \dot{p}_{1_-}(t)
                    &=
                   \frac{1}{4}\Big[ \gamma_-\Big(p_{1_+}(t)-3p_{1_-}(t)+p_{2_+}(t)+p_{2_-}(t)\Big)
                   \Big]
                    \\
                    \dot{p}_{2_+}(t)
                    &=
                    \frac{1}{4}\Big[
                    \gamma_+\,p_{1_+}(t)
                     +\gamma_-\,p_{1_-}(t)
                     -(2\gamma_++\gamma_-)\,p_{2_+}(t)
                     +\gamma_+\,p_{2_-}(t)
                    \Big]
                    \\
                    \dot{p}_{2_-}(t)
                    &=
                    \frac{1}{4}\Big[
                    \gamma_+\,p_{1_+}(t)
                     +\gamma_-\,p_{1_-}(t)
                     +\gamma_+\,p_{2_+}(t)
                     -(2\gamma_++\gamma_-)\,p_{2_-}(t)
                    \Big]
                \end{aligned}
                \right.
            \end{equation}
        \end{widetext}
        whose solution is given by
        \begin{equation}
	    \label{depsolution}
	        \left\{
	        \begin{aligned}
    	        p_{1_-}(t)
    	        &=p_{1_-}(\td)\,e^{-\gamma_-(t-\td)}
    	        +\frac{1}{4}\Big(1-e^{-\gamma_-(t-\td)}\Big)
    	        \\
    	        p_v(t)
    	        &=p_v(\td)\,e^{-(3\gamma_++\gamma_-)(t-\td)/4}
    	        +\frac{1}{4}\Big(1-e^{-(3\gamma_++\gamma_-)(t-\td)/4}\Big)
    	        \\&
    	        +\frac{1-4p_{1_-}(\td)}{12}\Big(e^{-\gamma_-(t-\td)}-e^{-(3\gamma_++\gamma_-)(t-\td)/4}\Big),
	        \end{aligned}
	        \right.
	    \end{equation}
	    with $v=1_+,\,2_+,\,2_-$. Eq.\tilde(\ref{wernerindist}) tells us that $p_{1_+}(\td)=p_{2_+}(\td)=p_{2_-}(\td)=\frac{1}{4}\,p(\td)$ and $p_{1_-}(\td)=1-\frac{3}{4}\,p(\td)$, so that evolving the system with Eq.\tilde(\ref{depsolution}) and performing the sLOCC measurement of Eq.\tilde(\ref{sloccstate}) we obtain the final normalized distinguishable state $\rlr(t)$ given in Eq.\tilde(\ref{nondissipativefinalstate}) with $p_{1_\pm}(t),\,p_{2_\pm}(t)$ as in Eq.\tilde(\ref{depsolution}).
	   
	    \begin{figure}[t!]
		    \centering
		    \includegraphics[width=0.49\textwidth]{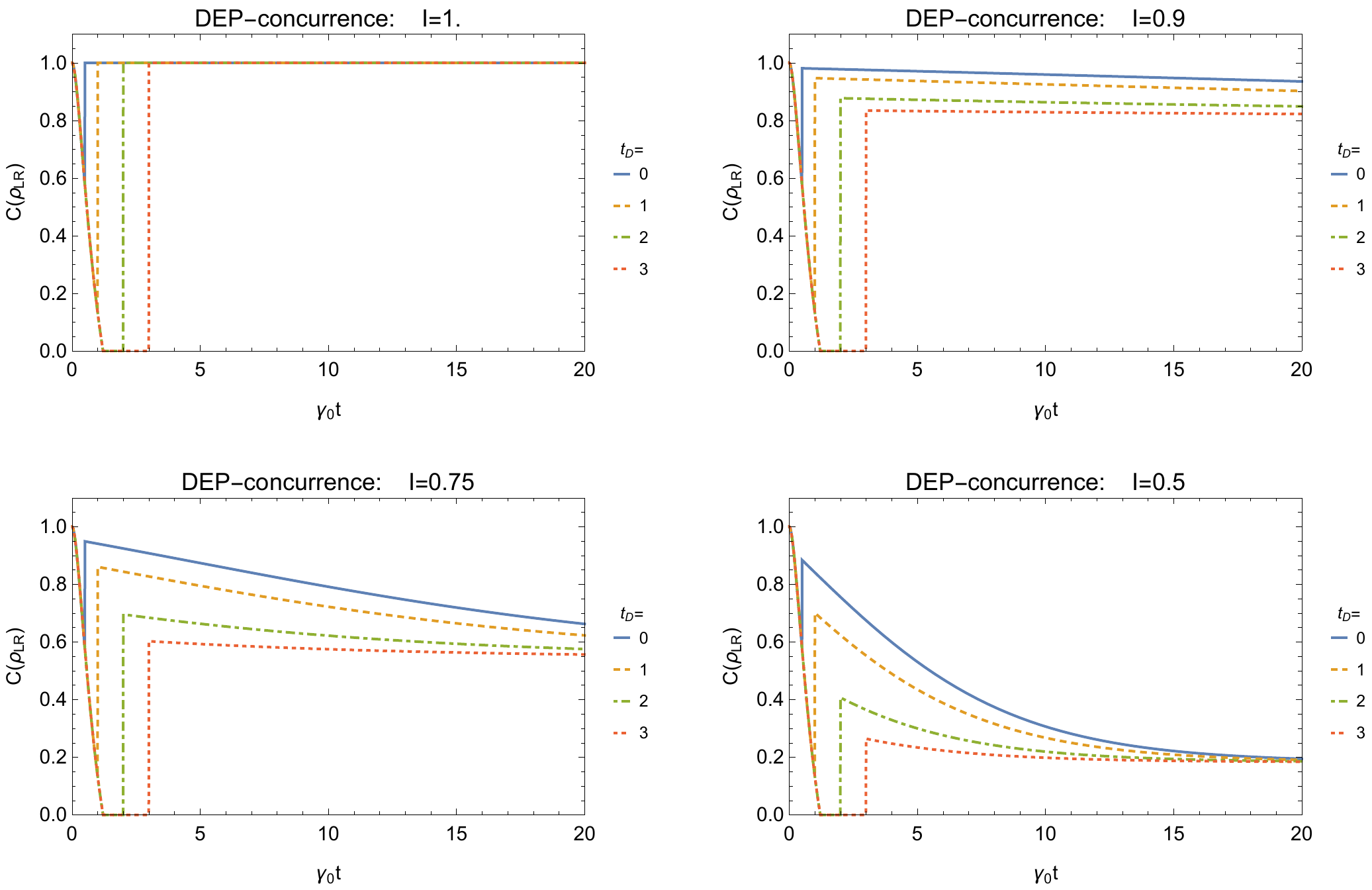}
		    \caption{\footnotesize Concurrence $C(\rho_\mathrm{LR})$ for the case of separated depolarizing channels as a function of the total dimensionless interaction time $\gamma_0 t$. Four different deformation times $\gamma_0 t_D$ for four values of spatial indistinguishability $\mathcal{I}$ are reported, under the constraint $|l'|=|r|$. Results hold for fermions with $l,r,l',$ and $r'$ positive and for bosons with one of such coefficients negative.}
		\label{depconc}
	    \end{figure}
	    
    In Figure\tilde\ref{depconc} we show the concurrence of such a state for different deformation times and different values of spatial indistinguishability. As can be noticed, the same general behaviour of the phase damping channel is found. In particular, our procedure allows the recovery of the initial quantum correlations spoiled by the detrimental action of the noisy environment. Such a restoration is partial when the spatial overlap between the two qubits wave functions is partial, and complete when the degree of indistinguishability $\mathcal{I}$ achieved is maximum. For every $\mathcal{I}<1$, the amount of entanglement restored increases as the interaction time decreases; nonetheless, the presence of an horizontal asymptote indicates that our procedure always allows for the recovery of a certain amount of correlations independently on the interaction time. Finally, such an amount is found to increase with the degree of spatial indistinguishability.

        \begin{figure}[t!]
		    \centering
		    \includegraphics[width=0.49\textwidth]{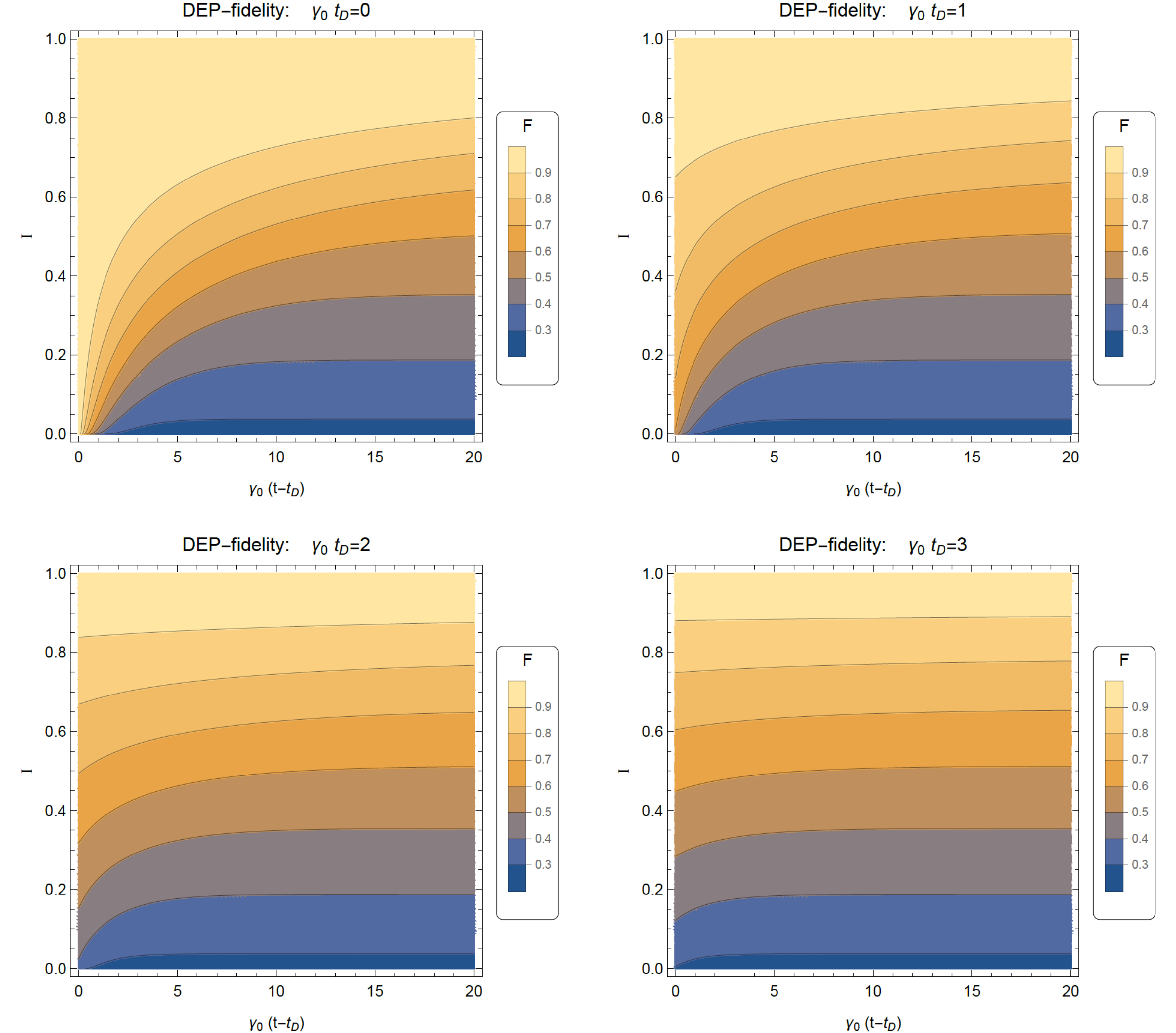}
		    \caption{\footnotesize Fidelity $F$ between final state and initial Bell singlet state in the depolarizing scenario versus dimensionless interaction time after the deformation $\gamma_0 (t-t_D)$ and $\mathcal{I}$, with constraint $|l'|=|r|$. Four different interaction times are considered. Results hold for fermions with $l,r,l',$ and $r'$ positive and for bosons with one of such coefficients negative.}
		\label{depfid}
	    \end{figure}
	     
	    \begin{figure}[t!]
		    \centering
		    \includegraphics[width=0.49\textwidth]{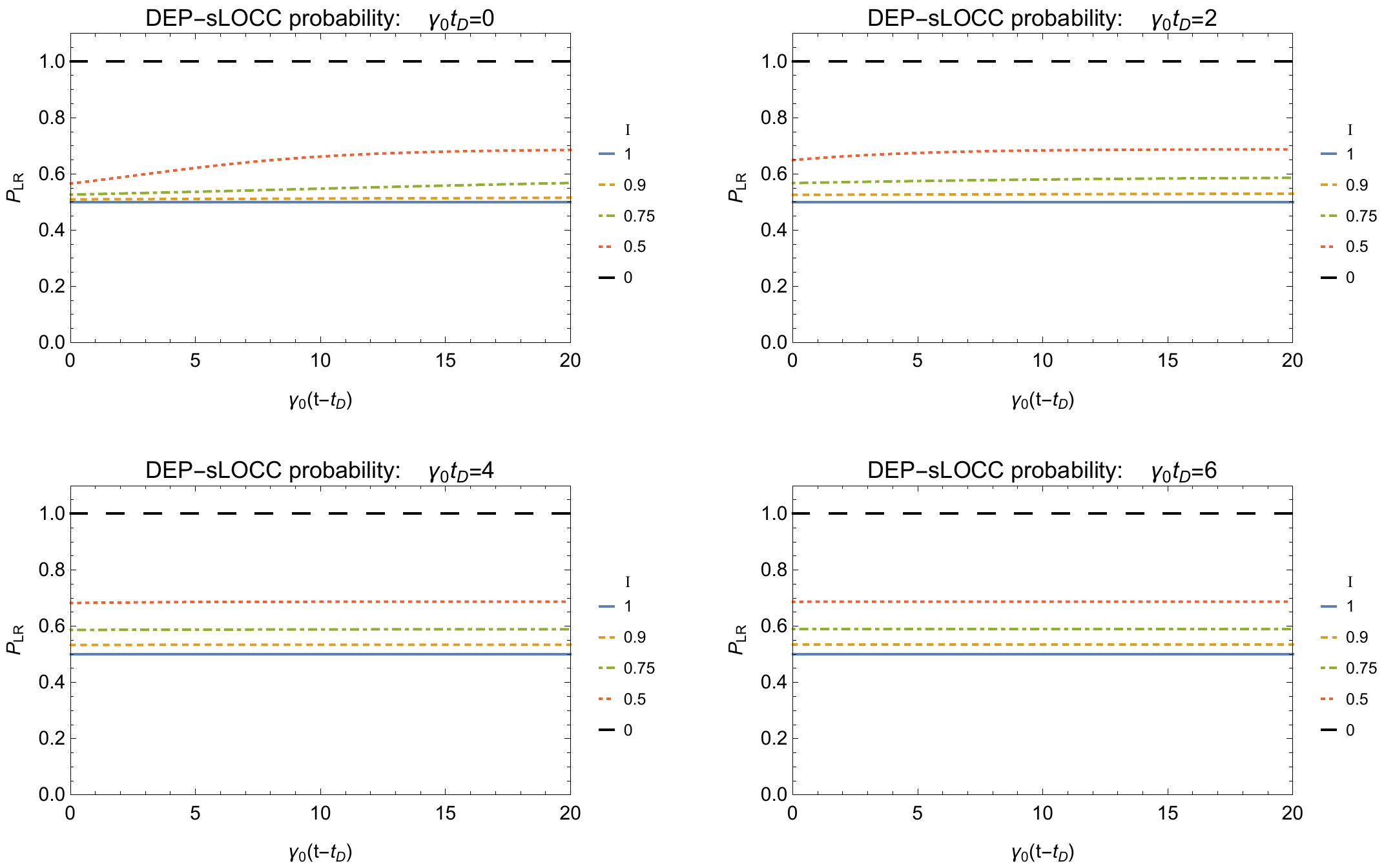}
		    \caption{\footnotesize Postselection (sLOCC) probability $P_\mathrm{LR}$ for the case of separated depolarizing channels as a function of the dimensionless interaction time after the deformation $\gamma_0 (t-t_D)$. Five different values of $\mathcal{I}$ for four different deformation times are reported, under the constraint $|l'|=|r|$. Results hold for fermions with $l,r,l',$ and $r'$ positive and for bosons with one of such coefficients negative.}
		\label{depprob}
	    \end{figure}
	       
        The similarity with the phase damping channels also holds when analyzing the fidelity of Eq.\tilde(\ref{fidelity}) between the final state of Eq.\tilde(\ref{nondissipativefinalstate}) and the initial $\kom$, plotted in Figure\tilde\ref{depfid} as a function of the dimensionless interaction time after the deformation and the degree of spatial indistinguishability achieved.
	    Here, the final state is shown to be closer to the initial one as the degree of spatial indistinguishability increases at fixed interaction times, and as the interaction time decreases at fixed $\mathcal{I}$. Furthermore, when $\mathcal{I}=1$ the fidelity is maximum independently on the duration of the interaction, indicating the complete entanglement restoration previously discussed.

	    Finally, we report in Figure\tilde\ref{depprob} the probability of success of the sLOCC postselection, which is once again given by Eq.\tilde(\ref{nondissipativeprob}) with the populations as in Eq.\tilde(\ref{depsolution}).
	  The same general behaviour of the phase damping channel is found, namely a trade-off between the concurrence and $P_\textrm{LR}$ resulting in the latest to decrease with the degree of spatial indistinguishability, starting from its maximum $P_\textrm{LR}=1$ when $\mathcal{I}=0$ and reaching its minimum $P_\textrm{LR}=0.5$ for distinguishable particles. An horizontal asymptote at an indistinguishability-dependent value comprised between $P_\textrm{LR}=0.5$ and $P_\textrm{LR}=1$ is reached from below as the interaction time grows when $0<\mathcal{I}<1$.
	  \\

	    \subsection{Amplitude damping channel}
        Finally, we study the amplitude damping channel scenario. Such a dissipative model describes a qubit interacting with an environment via the spontaneous emission of a quantum of excitation. The amplitude damping channel has found wide success in describing system such as atoms trapped in cavities\tilde\cite{bruzewicz,myatt,schindler}, single photons scattering in cavity QED\tilde\cite{nielsen2010quantum}, high temperature spin systems relaxing to the equilibrium state with their environment \cite{nielsen2010quantum}, superconducting qubits in circuit QED\tilde\cite{blais_2007,blais_2020}, and spin chains subjected to a ferromagnetic Heisenberg interaction\tilde\cite{giovannetti}.
        
        The Kraus operators for an amplitude damping channel acting on a single qubit are
        \begin{equation}
    	\label{krausadc}
        	\begin{gathered}
            	E_0=\sqrt{1-p(t)}\,\ku\bu+\kd\bd
            	=E_0^{\dagger},
            	\\
            	E_1=\sqrt{p(t)}\,\kd\bu,
            	\quad
            	E_1^{\dagger}=\sqrt{p(t)}\,\ku\bd.
        	\end{gathered}
    	\end{equation}
    	Starting from the pure state $\kom_\textrm{AB}$, the evolution given by Eq.\tilde(\ref{kraus}) with the operators above leads to
    	\begin{equation}
    	\label{distad}
        	\rab(t)
        	=\Big(1-p(t)\Big)\kom_\ab\bom_\ab
        	+p(t)\ket{D}_\textrm{AB}\bra{D}_\textrm{AB}.
    	\end{equation}

	    \begin{figure}[t!]
		    \centering
		    \includegraphics[width=0.49\textwidth]{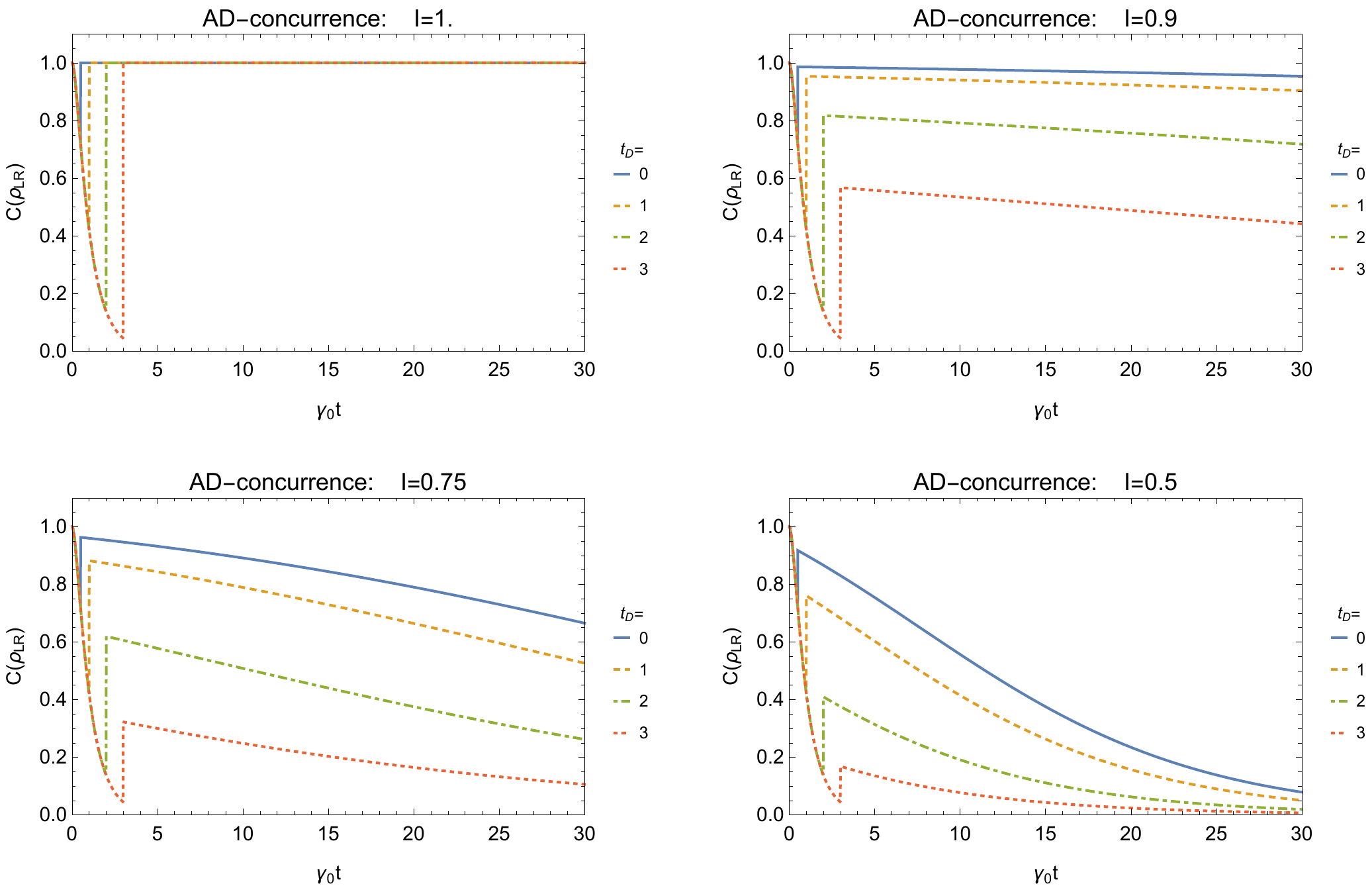}
		    \caption{\footnotesize Concurrence $C(\rho_\mathrm{LR})$ for the case of separated amplitude damping channels as a function of the total dimensionless interaction time $\gamma_0 t$. Four different deformation times $\gamma_0 t_D$ for four values of spatial indistinguishability $\mathcal{I}$ are reported, under the constraint $|l'|=|r|$. Results hold for fermions with $l,r,l',$ and $r'$ positive and for bosons with one of such coefficients negative.}
		\label{adconc}
	    \end{figure}

    	Deforming the state of Eq.\tilde(\ref{distad}) with the deformation operator of Eq.\tilde(\ref{deformation}) at time $\td$ we get
    	\begin{equation}
    	\label{indistad}
    	    \rho_\textrm{D}(\td)
    	    =\Big(1-p(\td)\Big)\dkom\dbom
        	+p(\td)\dkd\dbd.
    	\end{equation}
	    The interaction between the system and the environment is given by the Hamiltonian of Eq.\tilde(\ref{interactionhamilt2}) with
	    \begin{equation}
	    \label{interactionad}
	        H_{I,X}^{(j)}
	        =S_-^{(j)}
	        \otimes\sum_k \Big(\lvert\braket{X|\psi_j}\rvert\,g_{kX}\Big) a_{kX}^\dagger + h.c.,
	    \end{equation}
	    showing that the system operator $S_X$ is given by the usual lowering operator $S_-:=S_x-iS_y$ mapping $\ket{\uparrow}$ into $\ket{\downarrow}$ and $\ket{\downarrow}$ into 0.
	    The corresponding master equation of Eq.\tilde(\ref{indistmastereq}) is
    	    \begin{equation}
            \label{admastereq}
                \frac{d}{dt}\rho_\textrm{D}(t)
                =\sum_{X=L,R}\,\sum_{i,j=1,2}
                \gamma_X^{(i,j)}
                \left[
                    S_-^{(i)}\,\rho_S(t)\,S_+^{(j)}-\frac{1}{2}\big\{S_+^{(i)} S_-^{(j)},\,\rho_S (t)\big\}
                \right],
            \end{equation}
	    with $S_+=S_-^\dagger$ being the usual raising operator. Such an evolution preserves the diagonal structure of a state expressed on the basis $\mathcal{\bar{B}}_2$.
	    Writing
	    \begin{equation}
            \rd(t)=\sum_{u\in\mathcal{\bar{B}}_2}p_u(t)\,\ket{u}\bra{u},
        \end{equation}
        the above master equation reduces to the system of ODEs
        \begin{gather}
        \label{adode}
            \left\{
            \begin{gathered}
                \dot{p}_{1_+}(t)
                =\frac{\gamma_+}{2}\Big(p_U(t)-p_{1_+}(t)\Big)
                \\
                \dot{p}_{1_-}(t)
                =\frac{\gamma_-}{2}\Big(p_U(t)-p_{1_-}(t)\Big)
                \\
                \dot{p}_{U}(t)
                =-\frac{1}{2}\big(\gamma_++\gamma_-\big)\,p_U(t)
                \\
                \dot{p}_{D}(t)
                =\frac{1}{2}\Big(\gamma_+\,p_{1_+}(t)+\gamma_-\,p_{1_-}(t)\Big).
            \end{gathered}
            \right.
        \end{gather}
        By solving it, we get
            \begin{equation}
    	    \label{adsolution}
    	        \left\{
    	        \begin{aligned}
        	        p_{1_+}(t)
        	        &=p_{1_+}(\td)\,e^{-\gamma_+(t-\td)/2}
        	        \\&+
        	        p_U(\td)\left[\frac{\gamma_+}{\gamma_-}\left(e^{-\gamma_+(t-\td)/2}-e^{-2\gamma_0(t-\td)}\right)\right]
        	        \\
        	        p_{1_-}(t)
        	        &=p_{1_-}(\td)\,e^{-\gamma_-(t-\td)/2}
        	        \\&+
        	        p_U(\td)\left[\frac{\gamma_-}{\gamma_+}\left(e^{-\gamma_-(t-\td)/2}-e^{-2\gamma_0(t-\td)}\right)\right]
        	        \\
        	        p_U(t)
        	        &=p_U(\td)\,e^{-2\gamma_0(t-\td)}
        	        \\
        	        p_D(t)
        	        &=1-p_{1_+}(t)-p_{1_-}(t)-p_U(t).
    	        \end{aligned}
    	        \right.
    	    \end{equation}
	    Looking at Eq.\tilde(\ref{indistad}) we set $p_{1_-}=1-p(\td),\,p_D(t)=p(\td)$, and $p_{1_+}(t)=p_U(t)=0$ to evolve the system's state according to Eq.\tilde(\ref{adsolution}) until time $t$, when the sLOCC measurement of Eq.\tilde(\ref{sloccstate}) is performed. The result is the final distinguishable state
	    \begin{equation}
	    \label{adfinalstate}
            \begin{aligned}
	            \rlr(t)
	            =\frac{1}{N}\bigg[&\absp
	            \sum_{v=1_+,U,D}p_v(t)\ket{v}_\textrm{LR}\bra{v}_\textrm{LR}\\
	            +&\absm p_{1_-}(t)\kom_{\textrm{LR}}\bom_{\textrm{LR}}\bigg],
	        \end{aligned}
	    \end{equation}
	    with
	    \[
	        N:=\absp\sum_{v=1_+,U,D}p_v(t)+\absm p_{1_-}(t).
	    \]

	     The concurrence of the state of Eq.\tilde(\ref{adfinalstate}) is shown in Figure\tilde\ref{adconc} as a function of the total dimensionless interaction time for different degrees of spatial indistinguishability and deformation times.
	    As can be noticed, results show some common aspects with the two other types of noise considered; in particular, our procedure still clearly allows for the recovery of the spoiled quantum correlations in an amount which increases with the degree of spatial indistinguishability. Despite the similarities, the dissipative nature of the channel manifests itself in a sudden death phenomenon for the concurrence, showing that in this scenario the effectiveness of the described process is bound to the interaction time. Nonetheless, entanglement restoration is still possible on time scales increasing to infinity with $\mathcal{I}$ to the point of reaching a time independent-complete state protection for $\mathcal{I}=1$, thus preserving the validity of our procedure.
	    
	    \begin{figure}[t!]
		    \centering
		    \includegraphics[width=0.49\textwidth]{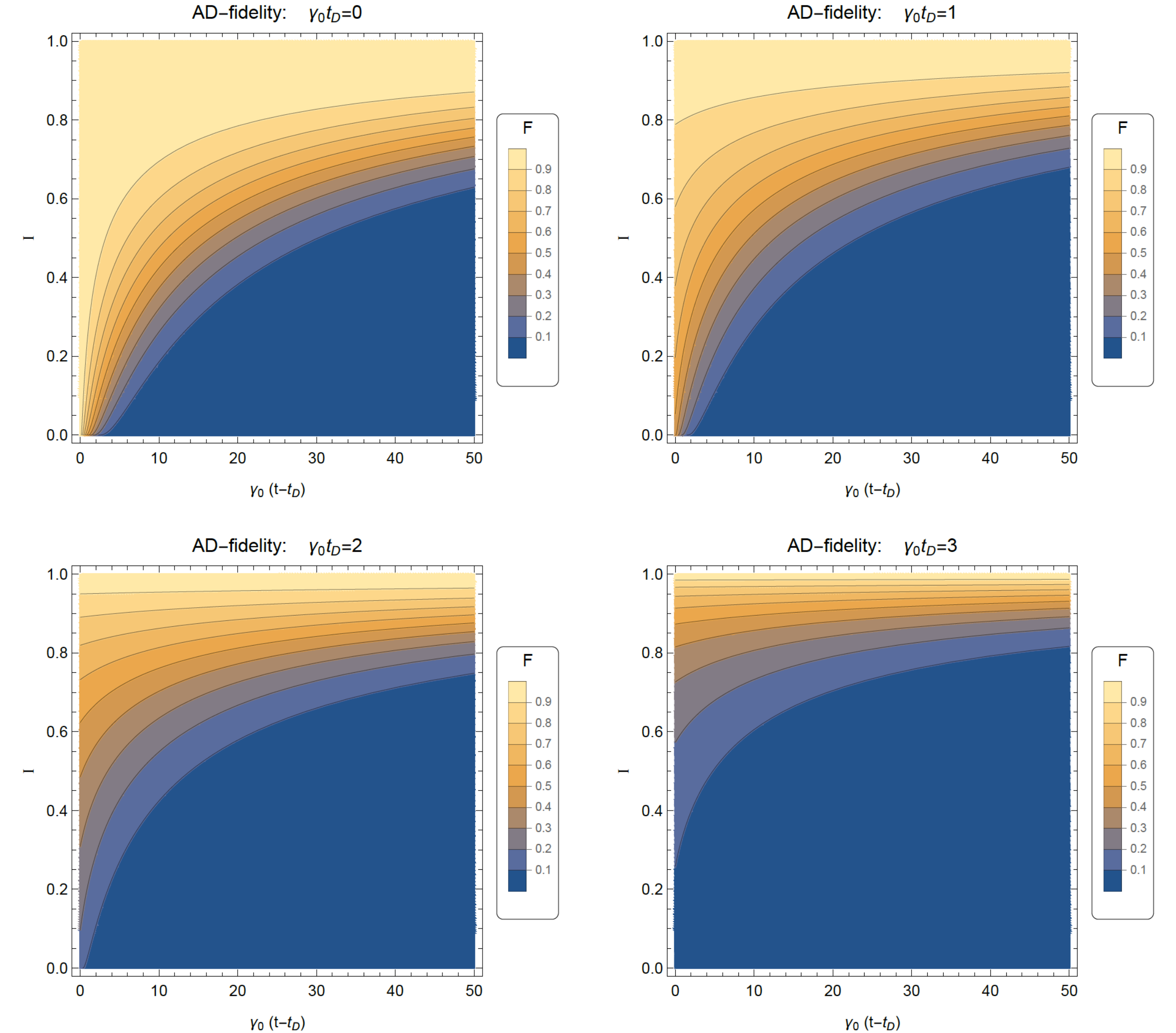}
		    \caption{\footnotesize Fidelity $F$ between final state and initial Bell singlet state in the amplitude damping scenario as a function of dimensionless interaction time after the deformation $\gamma_0 (t-t_D)$ and $\mathcal{I}$, with constraint $|l'|=|r|$. Four different interaction times are considered. Results hold for fermions with $l,r,l',$ and $r'$ positive and for bosons with one of such coefficients negative.}
		\label{adfid}
	    \end{figure}
	    
	    In Figure\tilde\ref{adfid} we report the fidelity of Eq.\tilde(\ref{fidelity}) between the final state of Eq.\tilde(\ref{adfinalstate}) and the initial $\kom$ state as a function of the dimensionless interaction time after the deformation and the degree of spatial indistinguishability.
	    The above described behaviour is reflected in the fidelity increasing with the degree of spatial indistinguishability at a fixed interaction time, reaching a value $F=1$ for $\mathcal{I}=1$ as we recover the initial state, and decreasing with $\gamma_0\,t$ at fixed $\mathcal{I}$.
	    
	        \begin{figure}[t!]
		    \centering
		    \includegraphics[width=0.49\textwidth]{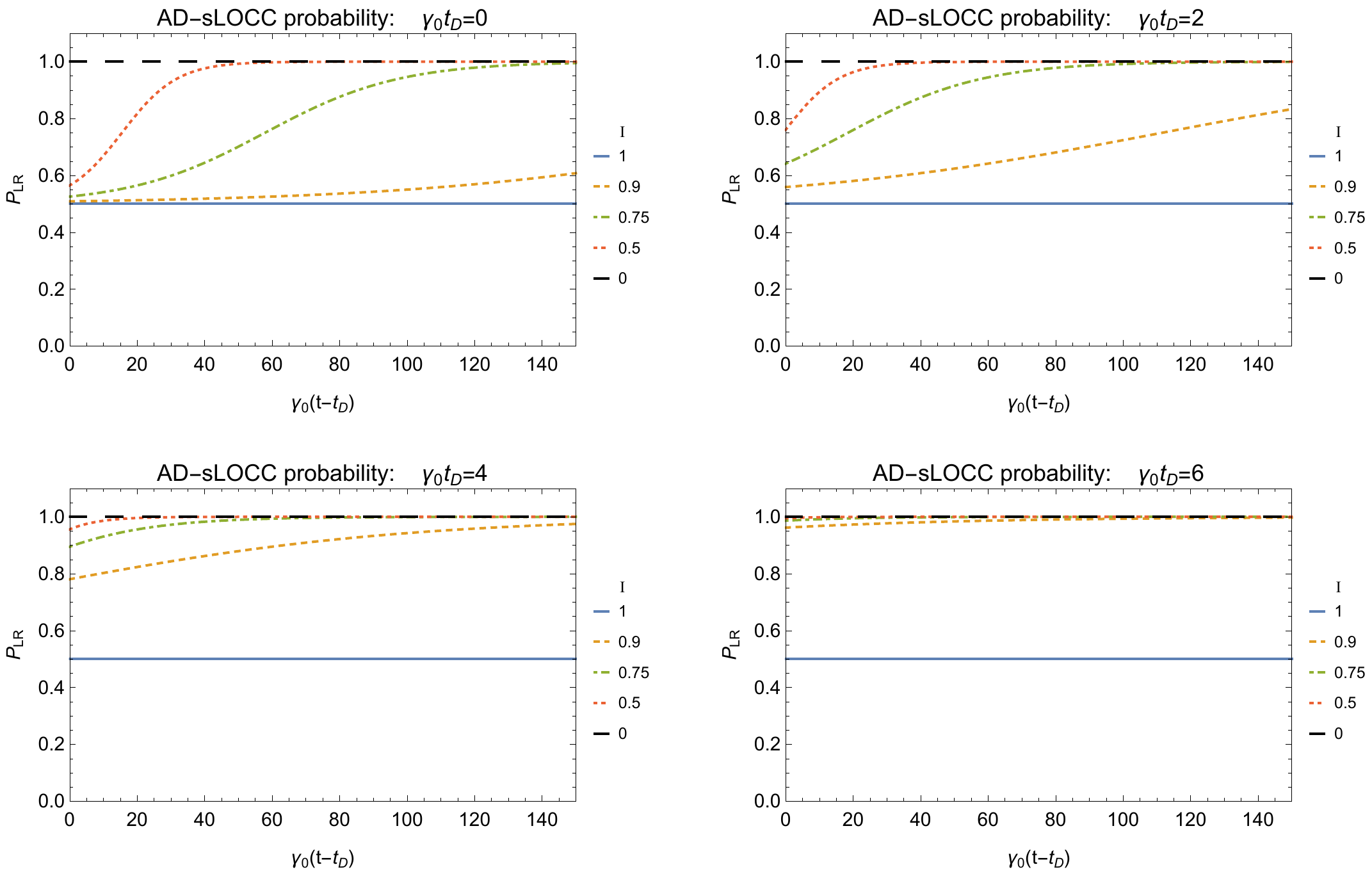}
		    \caption{\footnotesize Postselection (sLOCC) probability $P_\mathrm{LR}$ for the case of separated amplitude damping channels as a function of the dimensionless interaction time after the deformation $\gamma_0 (t-t_D)$.
		    Four different deformation times are reported. For each plot, five different values of $\mathcal{I}$ are considered, under the constraint $|l'|=|r|$. Results hold for fermions with $l,r,l',$ and $r'$ positive and for bosons with one of such coefficients negative.}
		\label{adprob}
	    \end{figure}
	    
	    To conclude this section, we report in Figure\tilde\ref{adprob} the probability of success of the sLOCC postselection.
	    The same trade-off is found for the other analyzed channels: the higher is the degree of indistinguishability and thus the entanglement restored, the more states are discarded. Once again, the minimum value for the probability is $P_\textrm{LR}=0.5$ when $\mathcal{I}=0$, while $P_\textrm{LR}=1$ for distinguishable particles. Nonetheless, this time the asymptotic value as the interaction time grows when $0<\mathcal{I}<1$ is always $P_\textrm{LR}=1$, as a byproduct of the sudden death phenomenon found for the concurrence.

   \section{Conclusions} 
    
    In this paper we have merged the results of Refs.\tilde\cite{indistdynamicalprotection} and\tilde\cite{Piccolini_2021} to propose a more general procedure capable of restoring the quantum correlations between two qubits undergoing the detrimental action of two independent noisy environments. Such a technique fundamentally relies on the spatial indistinguishability of identical constituents as a fundamental quantum resource made accessible by the sLOCC operational framework\tilde\cite{slocc}. Three standard types of noisy environments have been characterized in a Markovian regime: a phase damping channel, a depolarizing channel, and an amplitude damping channel, acting on the bipartite system both before and after the indistinguishability is generated via the deformation. Contextually, Refs.~\cite{indistdynamicalprotection} and \cite{Piccolini_2021} provide two limiting cases of this work, occurring respectively when the deformation is immediately performed after the generation of the initial Bell singlet state and when the sLOCC projective measurement is performed immediately after the deformation. The results of the two above mentioned papers are confirmed here even in the more general situation considered: quantum correlations are restored in an amount proportional to the degree of spatial indistinguishability achieved. Remarkably, when the spatial overlap is maximum (i.e. $\mathcal{I}=1$) the initial maximally entangled state is recovered, with a related maximum restoration of the entanglement. In all the analyzed situations, this comes with a trade-off given by a lowering of the sLOCC postselection probability $P_\textrm{LR}$, which never decreases below the value $P_\textrm{LR}=0.5$. Furthermore, in the two non-dissipative cases analyzed, our procedure has been found to be effective even for asymptotically large interaction times, while a sudden death phenomenon for the restored concurrence found in the amplitude damping scenario shows the urgency of performing the deformation and the subsequent sLOCC projection as soon as possible in a dissipative situation.

    \section*{Acknowledgments}
    
    The authors would like to acknowledge the importance of the work carried on by the late Prof.~Andrzej Kossakowski on the theory of open quantum systems: his long and prolific career has led, among many other impactful and significant results, to the development of the GKLS master equation used in this paper to determine the dynamics of the bipartite state populations for indistinguishable particles.
    
    R.M. thanks support from NSERC, MEI and the CRC program in Canada. R.L.F. thanks support from ``Sistema di Incentivazione, Sostegno e Premialit\`a della Ricerca Dipartimentale'' of the Department of Engineering, University of Palermo.


\begin{thebibliography}{96}
\expandafter\ifx\csname natexlab\endcsname\relax\def\natexlab#1{#1}\fi
\expandafter\ifx\csname bibnamefont\endcsname\relax
  \def\bibnamefont#1{#1}\fi
\expandafter\ifx\csname bibfnamefont\endcsname\relax
  \def\bibfnamefont#1{#1}\fi
\expandafter\ifx\csname citenamefont\endcsname\relax
  \def\citenamefont#1{#1}\fi
\expandafter\ifx\csname url\endcsname\relax
  \def\url#1{\texttt{#1}}\fi
\expandafter\ifx\csname urlprefix\endcsname\relax\def\urlprefix{URL }\fi
\providecommand{\bibinfo}[2]{#2}
\providecommand{\eprint}[2][]{\url{#2}}

\bibitem[{\citenamefont{Horodecki et~al.}(2009)\citenamefont{Horodecki,
  Horodecki, Horodecki, and Horodecki}}]{horodecki}
\bibinfo{author}{\bibfnamefont{R.}~\bibnamefont{Horodecki}},
  \bibinfo{author}{\bibfnamefont{P.}~\bibnamefont{Horodecki}},
  \bibinfo{author}{\bibfnamefont{M.}~\bibnamefont{Horodecki}},
  \bibnamefont{and}
  \bibinfo{author}{\bibfnamefont{K.}~\bibnamefont{Horodecki}},
  \bibinfo{journal}{Rev. Mod. Phys.} \textbf{\bibinfo{volume}{81}},
  \bibinfo{pages}{865} (\bibinfo{year}{2009}).

\bibitem[{\citenamefont{Ladd et~al.}(2010)\citenamefont{Ladd, Jelezko,
  Laflamme, Nakamura, Monroe, and O'Brien}}]{obrienreview}
\bibinfo{author}{\bibfnamefont{T.~D.} \bibnamefont{Ladd}},
  \bibinfo{author}{\bibfnamefont{F.}~\bibnamefont{Jelezko}},
  \bibinfo{author}{\bibfnamefont{R.}~\bibnamefont{Laflamme}},
  \bibinfo{author}{\bibfnamefont{Y.}~\bibnamefont{Nakamura}},
  \bibinfo{author}{\bibfnamefont{C.}~\bibnamefont{Monroe}}, \bibnamefont{and}
  \bibinfo{author}{\bibfnamefont{J.~L.} \bibnamefont{O'Brien}},
  \bibinfo{journal}{Nature} \textbf{\bibinfo{volume}{464}}, \bibinfo{pages}{45}
  (\bibinfo{year}{2010}).

\bibitem[{\citenamefont{Altman et~al.}(2021)}]{AltmanPRXQuantum}
\bibinfo{author}{\bibfnamefont{E.}~\bibnamefont{Altman}} \bibnamefont{et~al.},
  \bibinfo{journal}{PRX Quantum} \textbf{\bibinfo{volume}{2}},
  \bibinfo{pages}{017003} (\bibinfo{year}{2021}).

\bibitem[{\citenamefont{Monroe}(2002)}]{monroe2002quantum}
\bibinfo{author}{\bibfnamefont{C.}~\bibnamefont{Monroe}},
  \bibinfo{journal}{Nature} \textbf{\bibinfo{volume}{416}},
  \bibinfo{pages}{238} (\bibinfo{year}{2002}).

\bibitem[{\citenamefont{Pirandola et~al.}(2015)\citenamefont{Pirandola, Eisert,
  Weedbrook, Furusawa, and Braunstein}}]{pirandola2015advances}
\bibinfo{author}{\bibfnamefont{S.}~\bibnamefont{Pirandola}},
  \bibinfo{author}{\bibfnamefont{J.}~\bibnamefont{Eisert}},
  \bibinfo{author}{\bibfnamefont{C.}~\bibnamefont{Weedbrook}},
  \bibinfo{author}{\bibfnamefont{A.}~\bibnamefont{Furusawa}}, \bibnamefont{and}
  \bibinfo{author}{\bibfnamefont{S.~L.} \bibnamefont{Braunstein}},
  \bibinfo{journal}{Nature photonics} \textbf{\bibinfo{volume}{9}},
  \bibinfo{pages}{641} (\bibinfo{year}{2015}).

\bibitem[{\citenamefont{Breuer et~al.}(2002)\citenamefont{Breuer, Petruccione
  et~al.}}]{breuer2002theory}
\bibinfo{author}{\bibfnamefont{H.-P.} \bibnamefont{Breuer}},
  \bibinfo{author}{\bibfnamefont{F.}~\bibnamefont{Petruccione}},
  \bibnamefont{et~al.}, \emph{\bibinfo{title}{The theory of open quantum
  systems}} (\bibinfo{publisher}{Oxford University Press on Demand},
  \bibinfo{year}{2002}).

\bibitem[{\citenamefont{Suter and {\'A}lvarez}(2016)}]{suter2016colloquium}
\bibinfo{author}{\bibfnamefont{D.}~\bibnamefont{Suter}} \bibnamefont{and}
  \bibinfo{author}{\bibfnamefont{G.~A.} \bibnamefont{{\'A}lvarez}},
  \bibinfo{journal}{Rev. Mod. Phys.} \textbf{\bibinfo{volume}{88}},
  \bibinfo{pages}{041001} (\bibinfo{year}{2016}).

\bibitem[{\citenamefont{Preskill}(2018)}]{preskill_2018}
\bibinfo{author}{\bibfnamefont{J.}~\bibnamefont{Preskill}},
  \bibinfo{journal}{Quantum} \textbf{\bibinfo{volume}{2}}, \bibinfo{pages}{79}
  (\bibinfo{year}{2018}).

\bibitem[{\citenamefont{Rotter and Bird}(2015)}]{rotter_2015}
\bibinfo{author}{\bibfnamefont{I.}~\bibnamefont{Rotter}} \bibnamefont{and}
  \bibinfo{author}{\bibfnamefont{J.}~\bibnamefont{Bird}},
  \bibinfo{journal}{Rep. Prog. Phys.} \textbf{\bibinfo{volume}{78}},
  \bibinfo{pages}{114001} (\bibinfo{year}{2015}).

\bibitem[{\citenamefont{Preskill}(1998)}]{preskill_1998}
\bibinfo{author}{\bibfnamefont{J.}~\bibnamefont{Preskill}},
  \bibinfo{journal}{Proc. Math. Phys. Eng. Sci.}
  \textbf{\bibinfo{volume}{454}}, \bibinfo{pages}{385} (\bibinfo{year}{1998}).

\bibitem[{\citenamefont{Knill}(2005)}]{knill2005quantum}
\bibinfo{author}{\bibfnamefont{E.}~\bibnamefont{Knill}},
  \bibinfo{journal}{Nature} \textbf{\bibinfo{volume}{434}}, \bibinfo{pages}{39}
  (\bibinfo{year}{2005}).

\bibitem[{\citenamefont{Shor}(1995)}]{shor_1995}
\bibinfo{author}{\bibfnamefont{P.~W.} \bibnamefont{Shor}},
  \bibinfo{journal}{Phys. Rev. A} \textbf{\bibinfo{volume}{52}},
  \bibinfo{pages}{R2493} (\bibinfo{year}{1995}).

\bibitem[{\citenamefont{Steane}(1996)}]{steane_1996}
\bibinfo{author}{\bibfnamefont{A.~M.} \bibnamefont{Steane}},
  \bibinfo{journal}{Phys. Rev. Lett.} \textbf{\bibinfo{volume}{77}},
  \bibinfo{pages}{793} (\bibinfo{year}{1996}).

\bibitem[{\citenamefont{Zanardi and Rasetti}(1997)}]{zanardi1997noiseless}
\bibinfo{author}{\bibfnamefont{P.}~\bibnamefont{Zanardi}} \bibnamefont{and}
  \bibinfo{author}{\bibfnamefont{M.}~\bibnamefont{Rasetti}},
  \bibinfo{journal}{Phys. Rev. Lett.} \textbf{\bibinfo{volume}{79}},
  \bibinfo{pages}{3306} (\bibinfo{year}{1997}).

\bibitem[{\citenamefont{Lidar et~al.}(1998)\citenamefont{Lidar, Chuang, and
  Whaley}}]{lidar1998decoherence}
\bibinfo{author}{\bibfnamefont{D.~A.} \bibnamefont{Lidar}},
  \bibinfo{author}{\bibfnamefont{I.~L.} \bibnamefont{Chuang}},
  \bibnamefont{and} \bibinfo{author}{\bibfnamefont{K.~B.}
  \bibnamefont{Whaley}}, \bibinfo{journal}{Phys. Rev. Lett.}
  \textbf{\bibinfo{volume}{81}}, \bibinfo{pages}{2594} (\bibinfo{year}{1998}).

\bibitem[{\citenamefont{Viola and Lloyd}(1998)}]{Viola1998}
\bibinfo{author}{\bibfnamefont{L.}~\bibnamefont{Viola}} \bibnamefont{and}
  \bibinfo{author}{\bibfnamefont{S.}~\bibnamefont{Lloyd}},
  \bibinfo{journal}{Phys. Rev. A} \textbf{\bibinfo{volume}{58}},
  \bibinfo{pages}{2733} (\bibinfo{year}{1998}).

\bibitem[{\citenamefont{Viola and Knill}(2005)}]{viola2005random}
\bibinfo{author}{\bibfnamefont{L.}~\bibnamefont{Viola}} \bibnamefont{and}
  \bibinfo{author}{\bibfnamefont{E.}~\bibnamefont{Knill}},
  \bibinfo{journal}{Phys. Rev. Lett.} \textbf{\bibinfo{volume}{94}},
  \bibinfo{pages}{060502} (\bibinfo{year}{2005}).

\bibitem[{\citenamefont{D’Arrigo et~al.}(2014)\citenamefont{D’Arrigo,
  Lo~Franco, Benenti, Paladino, and Falci}}]{darrigo_2014_aop}
\bibinfo{author}{\bibfnamefont{A.}~\bibnamefont{D’Arrigo}},
  \bibinfo{author}{\bibfnamefont{R.}~\bibnamefont{Lo~Franco}},
  \bibinfo{author}{\bibfnamefont{G.}~\bibnamefont{Benenti}},
  \bibinfo{author}{\bibfnamefont{E.}~\bibnamefont{Paladino}}, \bibnamefont{and}
  \bibinfo{author}{\bibfnamefont{G.}~\bibnamefont{Falci}},
  \bibinfo{journal}{Ann. Phys.} \textbf{\bibinfo{volume}{350}},
  \bibinfo{pages}{211} (\bibinfo{year}{2014}).

\bibitem[{\citenamefont{Lo~Franco et~al.}(2014)\citenamefont{Lo~Franco,
  D'Arrigo, Falci, Compagno, and Paladino}}]{franco2014preserving}
\bibinfo{author}{\bibfnamefont{R.}~\bibnamefont{Lo~Franco}},
  \bibinfo{author}{\bibfnamefont{A.}~\bibnamefont{D'Arrigo}},
  \bibinfo{author}{\bibfnamefont{G.}~\bibnamefont{Falci}},
  \bibinfo{author}{\bibfnamefont{G.}~\bibnamefont{Compagno}}, \bibnamefont{and}
  \bibinfo{author}{\bibfnamefont{E.}~\bibnamefont{Paladino}},
  \bibinfo{journal}{Phys. Rev. B} \textbf{\bibinfo{volume}{90}},
  \bibinfo{pages}{054304} (\bibinfo{year}{2014}).

\bibitem[{\citenamefont{Orieux et~al.}(2015)\citenamefont{Orieux, D'Arrigo,
  Ferranti, Lo~Franco, Benenti, Paladino, Falci, Sciarrino, and
  Mataloni}}]{orieux_2015}
\bibinfo{author}{\bibfnamefont{A.}~\bibnamefont{Orieux}},
  \bibinfo{author}{\bibfnamefont{A.}~\bibnamefont{D'Arrigo}},
  \bibinfo{author}{\bibfnamefont{G.}~\bibnamefont{Ferranti}},
  \bibinfo{author}{\bibfnamefont{R.}~\bibnamefont{Lo~Franco}},
  \bibinfo{author}{\bibfnamefont{G.}~\bibnamefont{Benenti}},
  \bibinfo{author}{\bibfnamefont{E.}~\bibnamefont{Paladino}},
  \bibinfo{author}{\bibfnamefont{G.}~\bibnamefont{Falci}},
  \bibinfo{author}{\bibfnamefont{F.}~\bibnamefont{Sciarrino}},
  \bibnamefont{and} \bibinfo{author}{\bibfnamefont{P.}~\bibnamefont{Mataloni}},
  \bibinfo{journal}{Sci. Rep.} \textbf{\bibinfo{volume}{5}}, \bibinfo{pages}{1}
  (\bibinfo{year}{2015}).

\bibitem[{\citenamefont{Facchi et~al.}(2004)\citenamefont{Facchi, Lidar, and
  Pascazio}}]{facchi_2004}
\bibinfo{author}{\bibfnamefont{P.}~\bibnamefont{Facchi}},
  \bibinfo{author}{\bibfnamefont{D.}~\bibnamefont{Lidar}}, \bibnamefont{and}
  \bibinfo{author}{\bibfnamefont{S.}~\bibnamefont{Pascazio}},
  \bibinfo{journal}{Phys. Rev. A} \textbf{\bibinfo{volume}{69}},
  \bibinfo{pages}{032314} (\bibinfo{year}{2004}).

\bibitem[{\citenamefont{Lo~Franco
  et~al.}(2012{\natexlab{a}})\citenamefont{Lo~Franco, Bellomo, Andersson, and
  Compagno}}]{lo_franco_2012_pra}
\bibinfo{author}{\bibfnamefont{R.}~\bibnamefont{Lo~Franco}},
  \bibinfo{author}{\bibfnamefont{B.}~\bibnamefont{Bellomo}},
  \bibinfo{author}{\bibfnamefont{E.}~\bibnamefont{Andersson}},
  \bibnamefont{and} \bibinfo{author}{\bibfnamefont{G.}~\bibnamefont{Compagno}},
  \bibinfo{journal}{Phys. Rev. A} \textbf{\bibinfo{volume}{85}},
  \bibinfo{pages}{032318} (\bibinfo{year}{2012}{\natexlab{a}}).

\bibitem[{\citenamefont{Xu et~al.}(2013)\citenamefont{Xu, Sun, Li, Xu, Guo,
  Andersson, Lo~Franco, and Compagno}}]{xu_2013}
\bibinfo{author}{\bibfnamefont{J.-S.} \bibnamefont{Xu}},
  \bibinfo{author}{\bibfnamefont{K.}~\bibnamefont{Sun}},
  \bibinfo{author}{\bibfnamefont{C.-F.} \bibnamefont{Li}},
  \bibinfo{author}{\bibfnamefont{X.-Y.} \bibnamefont{Xu}},
  \bibinfo{author}{\bibfnamefont{G.-C.} \bibnamefont{Guo}},
  \bibinfo{author}{\bibfnamefont{E.}~\bibnamefont{Andersson}},
  \bibinfo{author}{\bibfnamefont{R.}~\bibnamefont{Lo~Franco}},
  \bibnamefont{and} \bibinfo{author}{\bibfnamefont{G.}~\bibnamefont{Compagno}},
  \bibinfo{journal}{Nat. Comm.} \textbf{\bibinfo{volume}{4}},
  \bibinfo{pages}{1} (\bibinfo{year}{2013}).

\bibitem[{\citenamefont{Damodarakurup et~al.}(2009)\citenamefont{Damodarakurup,
  Lucamarini, Di~Giuseppe, Vitali, and Tombesi}}]{damodarakurup_2009}
\bibinfo{author}{\bibfnamefont{S.}~\bibnamefont{Damodarakurup}},
  \bibinfo{author}{\bibfnamefont{M.}~\bibnamefont{Lucamarini}},
  \bibinfo{author}{\bibfnamefont{G.}~\bibnamefont{Di~Giuseppe}},
  \bibinfo{author}{\bibfnamefont{D.}~\bibnamefont{Vitali}}, \bibnamefont{and}
  \bibinfo{author}{\bibfnamefont{P.}~\bibnamefont{Tombesi}},
  \bibinfo{journal}{Phys. Rev. Lett.} \textbf{\bibinfo{volume}{103}},
  \bibinfo{pages}{040502} (\bibinfo{year}{2009}).

\bibitem[{\citenamefont{Cuevas et~al.}(2017)\citenamefont{Cuevas, Mari,
  De~Pasquale, Orieux, Massaro, Sciarrino, Mataloni, and
  Giovannetti}}]{cuevas_2017}
\bibinfo{author}{\bibfnamefont{{\'A}.}~\bibnamefont{Cuevas}},
  \bibinfo{author}{\bibfnamefont{A.}~\bibnamefont{Mari}},
  \bibinfo{author}{\bibfnamefont{A.}~\bibnamefont{De~Pasquale}},
  \bibinfo{author}{\bibfnamefont{A.}~\bibnamefont{Orieux}},
  \bibinfo{author}{\bibfnamefont{M.}~\bibnamefont{Massaro}},
  \bibinfo{author}{\bibfnamefont{F.}~\bibnamefont{Sciarrino}},
  \bibinfo{author}{\bibfnamefont{P.}~\bibnamefont{Mataloni}}, \bibnamefont{and}
  \bibinfo{author}{\bibfnamefont{V.}~\bibnamefont{Giovannetti}},
  \bibinfo{journal}{Phys. Rev. A} \textbf{\bibinfo{volume}{96}},
  \bibinfo{pages}{012314} (\bibinfo{year}{2017}).

\bibitem[{\citenamefont{Bennett et~al.}(1996)\citenamefont{Bennett, Brassard,
  Popescu, Schumacher, Smolin, and Wootters}}]{Bennett1996}
\bibinfo{author}{\bibfnamefont{C.~H.} \bibnamefont{Bennett}},
  \bibinfo{author}{\bibfnamefont{G.}~\bibnamefont{Brassard}},
  \bibinfo{author}{\bibfnamefont{S.}~\bibnamefont{Popescu}},
  \bibinfo{author}{\bibfnamefont{B.}~\bibnamefont{Schumacher}},
  \bibinfo{author}{\bibfnamefont{J.~A.} \bibnamefont{Smolin}},
  \bibnamefont{and} \bibinfo{author}{\bibfnamefont{W.~K.}
  \bibnamefont{Wootters}}, \bibinfo{journal}{Phys. Rev. Lett.}
  \textbf{\bibinfo{volume}{76}}, \bibinfo{pages}{722} (\bibinfo{year}{1996}).

\bibitem[{\citenamefont{Kwiat et~al.}(2001)\citenamefont{Kwiat, Barraza-Lopez,
  Stefanov, and Gisin}}]{kwiat_2001}
\bibinfo{author}{\bibfnamefont{P.~G.} \bibnamefont{Kwiat}},
  \bibinfo{author}{\bibfnamefont{S.}~\bibnamefont{Barraza-Lopez}},
  \bibinfo{author}{\bibfnamefont{A.}~\bibnamefont{Stefanov}}, \bibnamefont{and}
  \bibinfo{author}{\bibfnamefont{N.}~\bibnamefont{Gisin}},
  \bibinfo{journal}{Nature} \textbf{\bibinfo{volume}{409}},
  \bibinfo{pages}{1014} (\bibinfo{year}{2001}).

\bibitem[{\citenamefont{Dong et~al.}(2008)\citenamefont{Dong, Lassen, Heersink,
  Marquardt, Filip, Leuchs, and Andersen}}]{dong_2008}
\bibinfo{author}{\bibfnamefont{R.}~\bibnamefont{Dong}},
  \bibinfo{author}{\bibfnamefont{M.}~\bibnamefont{Lassen}},
  \bibinfo{author}{\bibfnamefont{J.}~\bibnamefont{Heersink}},
  \bibinfo{author}{\bibfnamefont{C.}~\bibnamefont{Marquardt}},
  \bibinfo{author}{\bibfnamefont{R.}~\bibnamefont{Filip}},
  \bibinfo{author}{\bibfnamefont{G.}~\bibnamefont{Leuchs}}, \bibnamefont{and}
  \bibinfo{author}{\bibfnamefont{U.~L.} \bibnamefont{Andersen}},
  \bibinfo{journal}{Nat. Phys.} \textbf{\bibinfo{volume}{4}},
  \bibinfo{pages}{919} (\bibinfo{year}{2008}).

\bibitem[{\citenamefont{Mazzola et~al.}(2009)\citenamefont{Mazzola, Maniscalco,
  Piilo, Suominen, and Garraway}}]{mazzola_2009}
\bibinfo{author}{\bibfnamefont{L.}~\bibnamefont{Mazzola}},
  \bibinfo{author}{\bibfnamefont{S.}~\bibnamefont{Maniscalco}},
  \bibinfo{author}{\bibfnamefont{J.}~\bibnamefont{Piilo}},
  \bibinfo{author}{\bibfnamefont{K.-A.} \bibnamefont{Suominen}},
  \bibnamefont{and} \bibinfo{author}{\bibfnamefont{B.~M.}
  \bibnamefont{Garraway}}, \bibinfo{journal}{Phys. Rev. A}
  \textbf{\bibinfo{volume}{79}}, \bibinfo{pages}{042302}
  (\bibinfo{year}{2009}).

\bibitem[{\citenamefont{Bellomo et~al.}(2008)\citenamefont{Bellomo, Lo~Franco,
  Maniscalco, and Compagno}}]{bellomo_2008}
\bibinfo{author}{\bibfnamefont{B.}~\bibnamefont{Bellomo}},
  \bibinfo{author}{\bibfnamefont{R.}~\bibnamefont{Lo~Franco}},
  \bibinfo{author}{\bibfnamefont{S.}~\bibnamefont{Maniscalco}},
  \bibnamefont{and} \bibinfo{author}{\bibfnamefont{G.}~\bibnamefont{Compagno}},
  \bibinfo{journal}{Phys. Rev. A} \textbf{\bibinfo{volume}{78}},
  \bibinfo{pages}{060302} (\bibinfo{year}{2008}).

\bibitem[{\citenamefont{Lo~Franco et~al.}(2013)\citenamefont{Lo~Franco,
  Bellomo, Maniscalco, and Compagno}}]{lo_franco_2013}
\bibinfo{author}{\bibfnamefont{R.}~\bibnamefont{Lo~Franco}},
  \bibinfo{author}{\bibfnamefont{B.}~\bibnamefont{Bellomo}},
  \bibinfo{author}{\bibfnamefont{S.}~\bibnamefont{Maniscalco}},
  \bibnamefont{and} \bibinfo{author}{\bibfnamefont{G.}~\bibnamefont{Compagno}},
  \bibinfo{journal}{International Journal of Modern Physics B}
  \textbf{\bibinfo{volume}{27}}, \bibinfo{pages}{1345053}
  (\bibinfo{year}{2013}).

\bibitem[{\citenamefont{Aolita et~al.}(2015)\citenamefont{Aolita, De~Melo, and
  Davidovich}}]{aolita_2015}
\bibinfo{author}{\bibfnamefont{L.}~\bibnamefont{Aolita}},
  \bibinfo{author}{\bibfnamefont{F.}~\bibnamefont{De~Melo}}, \bibnamefont{and}
  \bibinfo{author}{\bibfnamefont{L.}~\bibnamefont{Davidovich}},
  \bibinfo{journal}{Rep. Prog. Phys.} \textbf{\bibinfo{volume}{78}},
  \bibinfo{pages}{042001} (\bibinfo{year}{2015}).

\bibitem[{\citenamefont{Xu et~al.}(2010)\citenamefont{Xu, Li, Gong, Zou, Shi,
  Chen, and Guo}}]{xu_2010}
\bibinfo{author}{\bibfnamefont{J.-S.} \bibnamefont{Xu}},
  \bibinfo{author}{\bibfnamefont{C.-F.} \bibnamefont{Li}},
  \bibinfo{author}{\bibfnamefont{M.}~\bibnamefont{Gong}},
  \bibinfo{author}{\bibfnamefont{X.-B.} \bibnamefont{Zou}},
  \bibinfo{author}{\bibfnamefont{C.-H.} \bibnamefont{Shi}},
  \bibinfo{author}{\bibfnamefont{G.}~\bibnamefont{Chen}}, \bibnamefont{and}
  \bibinfo{author}{\bibfnamefont{G.-C.} \bibnamefont{Guo}},
  \bibinfo{journal}{Phys. Rev. Lett.} \textbf{\bibinfo{volume}{104}},
  \bibinfo{pages}{100502} (\bibinfo{year}{2010}).

\bibitem[{\citenamefont{Bylicka et~al.}(2014)\citenamefont{Bylicka,
  Chru{\'s}ci{\'n}ski, and Maniscalco}}]{bylicka_2014}
\bibinfo{author}{\bibfnamefont{B.}~\bibnamefont{Bylicka}},
  \bibinfo{author}{\bibfnamefont{D.}~\bibnamefont{Chru{\'s}ci{\'n}ski}},
  \bibnamefont{and}
  \bibinfo{author}{\bibfnamefont{S.}~\bibnamefont{Maniscalco}},
  \bibinfo{journal}{Sci. Rep.} \textbf{\bibinfo{volume}{4}}, \bibinfo{pages}{1}
  (\bibinfo{year}{2014}).

\bibitem[{\citenamefont{Man et~al.}(2015{\natexlab{a}})\citenamefont{Man, Xia,
  and Franco}}]{man_2015}
\bibinfo{author}{\bibfnamefont{Z.-X.} \bibnamefont{Man}},
  \bibinfo{author}{\bibfnamefont{Y.-J.} \bibnamefont{Xia}}, \bibnamefont{and}
  \bibinfo{author}{\bibfnamefont{R.~L.} \bibnamefont{Franco}},
  \bibinfo{journal}{Sci. Rep.} \textbf{\bibinfo{volume}{5}}, \bibinfo{pages}{1}
  (\bibinfo{year}{2015}{\natexlab{a}}).

\bibitem[{\citenamefont{Tan et~al.}(2010)\citenamefont{Tan, Kyaw, and
  Yeo}}]{tan_2010}
\bibinfo{author}{\bibfnamefont{J.}~\bibnamefont{Tan}},
  \bibinfo{author}{\bibfnamefont{T.~H.} \bibnamefont{Kyaw}}, \bibnamefont{and}
  \bibinfo{author}{\bibfnamefont{Y.}~\bibnamefont{Yeo}},
  \bibinfo{journal}{Phys. Rev. A} \textbf{\bibinfo{volume}{81}},
  \bibinfo{pages}{062119} (\bibinfo{year}{2010}).

\bibitem[{\citenamefont{Tong et~al.}(2010)\citenamefont{Tong, An, Luo, and
  Oh}}]{tong_2010}
\bibinfo{author}{\bibfnamefont{Q.-J.} \bibnamefont{Tong}},
  \bibinfo{author}{\bibfnamefont{J.-H.} \bibnamefont{An}},
  \bibinfo{author}{\bibfnamefont{H.-G.} \bibnamefont{Luo}}, \bibnamefont{and}
  \bibinfo{author}{\bibfnamefont{C.}~\bibnamefont{Oh}}, \bibinfo{journal}{Phys.
  Rev. A} \textbf{\bibinfo{volume}{81}}, \bibinfo{pages}{052330}
  (\bibinfo{year}{2010}).

\bibitem[{\citenamefont{Breuer et~al.}(2016)\citenamefont{Breuer, Laine, Piilo,
  and Vacchini}}]{breuer_2016_colloquium}
\bibinfo{author}{\bibfnamefont{H.-P.} \bibnamefont{Breuer}},
  \bibinfo{author}{\bibfnamefont{E.-M.} \bibnamefont{Laine}},
  \bibinfo{author}{\bibfnamefont{J.}~\bibnamefont{Piilo}}, \bibnamefont{and}
  \bibinfo{author}{\bibfnamefont{B.}~\bibnamefont{Vacchini}},
  \bibinfo{journal}{Rev. Mod. Phys.} \textbf{\bibinfo{volume}{88}},
  \bibinfo{pages}{021002} (\bibinfo{year}{2016}).

\bibitem[{\citenamefont{Man et~al.}(2015{\natexlab{b}})\citenamefont{Man, Xia,
  and Lo~Franco}}]{man_2015_pra}
\bibinfo{author}{\bibfnamefont{Z.-X.} \bibnamefont{Man}},
  \bibinfo{author}{\bibfnamefont{Y.-J.} \bibnamefont{Xia}}, \bibnamefont{and}
  \bibinfo{author}{\bibfnamefont{R.}~\bibnamefont{Lo~Franco}},
  \bibinfo{journal}{Phys. Rev. A} \textbf{\bibinfo{volume}{92}},
  \bibinfo{pages}{012315} (\bibinfo{year}{2015}{\natexlab{b}}).

\bibitem[{\citenamefont{Ghirardi et~al.}(1977)\citenamefont{Ghirardi, Rimini,
  Weber, and Omero}}]{ghirardi1977some}
\bibinfo{author}{\bibfnamefont{G.~C.} \bibnamefont{Ghirardi}},
  \bibinfo{author}{\bibfnamefont{A.}~\bibnamefont{Rimini}},
  \bibinfo{author}{\bibfnamefont{T.}~\bibnamefont{Weber}}, \bibnamefont{and}
  \bibinfo{author}{\bibfnamefont{C.}~\bibnamefont{Omero}}, \bibinfo{journal}{Il
  Nuovo Cimento B (1971-1996)} \textbf{\bibinfo{volume}{39}},
  \bibinfo{pages}{130} (\bibinfo{year}{1977}).

\bibitem[{\citenamefont{Ghirardi and Marinatto}(2004)}]{ghirardi}
\bibinfo{author}{\bibfnamefont{G.}~\bibnamefont{Ghirardi}} \bibnamefont{and}
  \bibinfo{author}{\bibfnamefont{L.}~\bibnamefont{Marinatto}},
  \bibinfo{journal}{Phys. Rev. A} \textbf{\bibinfo{volume}{70}},
  \bibinfo{pages}{012109} (\bibinfo{year}{2004}).

\bibitem[{\citenamefont{Shi}(2003)}]{shi2003quantum}
\bibinfo{author}{\bibfnamefont{Y.}~\bibnamefont{Shi}},
  \bibinfo{journal}{Physical Review A} \textbf{\bibinfo{volume}{67}},
  \bibinfo{pages}{024301} (\bibinfo{year}{2003}).

\bibitem[{\citenamefont{Sasaki et~al.}(2011{\natexlab{a}})\citenamefont{Sasaki,
  Ichikawa, and Tsutsui}}]{sasaki2011entanglement}
\bibinfo{author}{\bibfnamefont{T.}~\bibnamefont{Sasaki}},
  \bibinfo{author}{\bibfnamefont{T.}~\bibnamefont{Ichikawa}}, \bibnamefont{and}
  \bibinfo{author}{\bibfnamefont{I.}~\bibnamefont{Tsutsui}},
  \bibinfo{journal}{Physical Review A} \textbf{\bibinfo{volume}{83}},
  \bibinfo{pages}{012113} (\bibinfo{year}{2011}{\natexlab{a}}).

\bibitem[{\citenamefont{Morris et~al.}(2020{\natexlab{a}})\citenamefont{Morris,
  Yadin, Fadel, Zibold, Treutlein, and Adesso}}]{morris2020entanglement}
\bibinfo{author}{\bibfnamefont{B.}~\bibnamefont{Morris}},
  \bibinfo{author}{\bibfnamefont{B.}~\bibnamefont{Yadin}},
  \bibinfo{author}{\bibfnamefont{M.}~\bibnamefont{Fadel}},
  \bibinfo{author}{\bibfnamefont{T.}~\bibnamefont{Zibold}},
  \bibinfo{author}{\bibfnamefont{P.}~\bibnamefont{Treutlein}},
  \bibnamefont{and} \bibinfo{author}{\bibfnamefont{G.}~\bibnamefont{Adesso}},
  \bibinfo{journal}{Physical Review X} \textbf{\bibinfo{volume}{10}},
  \bibinfo{pages}{041012} (\bibinfo{year}{2020}{\natexlab{a}}).

\bibitem[{\citenamefont{Benatti et~al.}(2020)\citenamefont{Benatti, Floreanini,
  Franchini, and Marzolino}}]{benatti2020entanglement}
\bibinfo{author}{\bibfnamefont{F.}~\bibnamefont{Benatti}},
  \bibinfo{author}{\bibfnamefont{R.}~\bibnamefont{Floreanini}},
  \bibinfo{author}{\bibfnamefont{F.}~\bibnamefont{Franchini}},
  \bibnamefont{and}
  \bibinfo{author}{\bibfnamefont{U.}~\bibnamefont{Marzolino}},
  \bibinfo{journal}{Physics Reports}  (\bibinfo{year}{2020}).

\bibitem[{\citenamefont{Cunden et~al.}(2014)\citenamefont{Cunden, {Di Martino},
  Facchi, and Florio}}]{facchiIJQI}
\bibinfo{author}{\bibfnamefont{F.~D.} \bibnamefont{Cunden}},
  \bibinfo{author}{\bibfnamefont{S.}~\bibnamefont{{Di Martino}}},
  \bibinfo{author}{\bibfnamefont{P.}~\bibnamefont{Facchi}}, \bibnamefont{and}
  \bibinfo{author}{\bibfnamefont{G.}~\bibnamefont{Florio}},
  \bibinfo{journal}{Int. J. Quantum Inform.} \textbf{\bibinfo{volume}{12}},
  \bibinfo{pages}{461001} (\bibinfo{year}{2014}).

\bibitem[{\citenamefont{Li et~al.}(2001)\citenamefont{Li, Zeng, Liu, and
  Long}}]{Li2001PRA}
\bibinfo{author}{\bibfnamefont{Y.-S.} \bibnamefont{Li}},
  \bibinfo{author}{\bibfnamefont{B.}~\bibnamefont{Zeng}},
  \bibinfo{author}{\bibfnamefont{X.-S.} \bibnamefont{Liu}}, \bibnamefont{and}
  \bibinfo{author}{\bibfnamefont{G.-L.} \bibnamefont{Long}},
  \bibinfo{journal}{Phys. Rev. A} \textbf{\bibinfo{volume}{64}},
  \bibinfo{pages}{054302} (\bibinfo{year}{2001}).

\bibitem[{\citenamefont{Paskauskas and You}(2001)}]{Paskauskas2001PRA}
\bibinfo{author}{\bibfnamefont{R.}~\bibnamefont{Paskauskas}} \bibnamefont{and}
  \bibinfo{author}{\bibfnamefont{L.}~\bibnamefont{You}},
  \bibinfo{journal}{Phys. Rev. A} \textbf{\bibinfo{volume}{64}},
  \bibinfo{pages}{042310} (\bibinfo{year}{2001}).

\bibitem[{\citenamefont{Schliemann et~al.}(2001)\citenamefont{Schliemann,
  Cirac, Kus, Lewenstein, and Loss}}]{cirac2001PRA}
\bibinfo{author}{\bibfnamefont{J.}~\bibnamefont{Schliemann}},
  \bibinfo{author}{\bibfnamefont{J.~I.} \bibnamefont{Cirac}},
  \bibinfo{author}{\bibfnamefont{M.}~\bibnamefont{Kus}},
  \bibinfo{author}{\bibfnamefont{M.}~\bibnamefont{Lewenstein}},
  \bibnamefont{and} \bibinfo{author}{\bibfnamefont{D.}~\bibnamefont{Loss}},
  \bibinfo{journal}{Phys. Rev. A} \textbf{\bibinfo{volume}{64}},
  \bibinfo{pages}{022303} (\bibinfo{year}{2001}).

\bibitem[{\citenamefont{Zanardi}(2002)}]{zanardiPRA}
\bibinfo{author}{\bibfnamefont{P.}~\bibnamefont{Zanardi}},
  \bibinfo{journal}{Phys. Rev. A} \textbf{\bibinfo{volume}{65}},
  \bibinfo{pages}{042101} (\bibinfo{year}{2002}).

\bibitem[{\citenamefont{Eckert et~al.}(2002)\citenamefont{Eckert, Schliemann,
  Bruss, and Lewenstein}}]{eckert2002AnnPhys}
\bibinfo{author}{\bibfnamefont{K.}~\bibnamefont{Eckert}},
  \bibinfo{author}{\bibfnamefont{J.}~\bibnamefont{Schliemann}},
  \bibinfo{author}{\bibfnamefont{D.}~\bibnamefont{Bruss}}, \bibnamefont{and}
  \bibinfo{author}{\bibfnamefont{M.}~\bibnamefont{Lewenstein}},
  \bibinfo{journal}{Ann. Phys.} \textbf{\bibinfo{volume}{299}},
  \bibinfo{pages}{88} (\bibinfo{year}{2002}).

\bibitem[{\citenamefont{Balachandran et~al.}(2013)\citenamefont{Balachandran,
  Govindarajan, de~Queiroz, and Reyes-Lega}}]{balachandranPRL}
\bibinfo{author}{\bibfnamefont{A.~P.} \bibnamefont{Balachandran}},
  \bibinfo{author}{\bibfnamefont{T.~R.} \bibnamefont{Govindarajan}},
  \bibinfo{author}{\bibfnamefont{A.~R.} \bibnamefont{de~Queiroz}},
  \bibnamefont{and} \bibinfo{author}{\bibfnamefont{A.~F.}
  \bibnamefont{Reyes-Lega}}, \bibinfo{journal}{Phys. Rev. Lett.}
  \textbf{\bibinfo{volume}{110}}, \bibinfo{pages}{080503}
  (\bibinfo{year}{2013}).

\bibitem[{\citenamefont{Sasaki et~al.}(2011{\natexlab{b}})\citenamefont{Sasaki,
  Ichikawa, and Tsutsui}}]{sasaki2011PRA}
\bibinfo{author}{\bibfnamefont{T.}~\bibnamefont{Sasaki}},
  \bibinfo{author}{\bibfnamefont{T.}~\bibnamefont{Ichikawa}}, \bibnamefont{and}
  \bibinfo{author}{\bibfnamefont{I.}~\bibnamefont{Tsutsui}},
  \bibinfo{journal}{Phys. Rev. A} \textbf{\bibinfo{volume}{83}},
  \bibinfo{pages}{012113} (\bibinfo{year}{2011}{\natexlab{b}}).

\bibitem[{\citenamefont{Benenti et~al.}(2013)\citenamefont{Benenti, Siccardi,
  and Strini}}]{giulianoEPJD}
\bibinfo{author}{\bibfnamefont{G.}~\bibnamefont{Benenti}},
  \bibinfo{author}{\bibfnamefont{S.}~\bibnamefont{Siccardi}}, \bibnamefont{and}
  \bibinfo{author}{\bibfnamefont{G.}~\bibnamefont{Strini}},
  \bibinfo{journal}{Eur. Phys. J. D} \textbf{\bibinfo{volume}{67}},
  \bibinfo{pages}{83} (\bibinfo{year}{2013}).

\bibitem[{\citenamefont{Bose and Home}(2002)}]{bose2002indisting}
\bibinfo{author}{\bibfnamefont{S.}~\bibnamefont{Bose}} \bibnamefont{and}
  \bibinfo{author}{\bibfnamefont{D.}~\bibnamefont{Home}},
  \bibinfo{journal}{Phys. Rev. Lett.} \textbf{\bibinfo{volume}{88}},
  \bibinfo{pages}{050401} (\bibinfo{year}{2002}).

\bibitem[{\citenamefont{Bose and Home}(2013)}]{bose2013}
\bibinfo{author}{\bibfnamefont{S.}~\bibnamefont{Bose}} \bibnamefont{and}
  \bibinfo{author}{\bibfnamefont{D.}~\bibnamefont{Home}},
  \bibinfo{journal}{Phys. Rev. Lett.} \textbf{\bibinfo{volume}{110}},
  \bibinfo{pages}{140404} (\bibinfo{year}{2013}).

\bibitem[{\citenamefont{Tichy et~al.}(2013)\citenamefont{Tichy, {de Melo}, Kus,
  Mintert, and Buchleitner}}]{tichyFort}
\bibinfo{author}{\bibfnamefont{M.~C.} \bibnamefont{Tichy}},
  \bibinfo{author}{\bibfnamefont{F.}~\bibnamefont{{de Melo}}},
  \bibinfo{author}{\bibfnamefont{M.}~\bibnamefont{Kus}},
  \bibinfo{author}{\bibfnamefont{F.}~\bibnamefont{Mintert}}, \bibnamefont{and}
  \bibinfo{author}{\bibfnamefont{A.}~\bibnamefont{Buchleitner}},
  \bibinfo{journal}{Fortschr. Phys.} \textbf{\bibinfo{volume}{61}},
  \bibinfo{pages}{225} (\bibinfo{year}{2013}).

\bibitem[{\citenamefont{Killoran et~al.}(2014)\citenamefont{Killoran, Cramer,
  and Plenio}}]{PlenioExtracting}
\bibinfo{author}{\bibfnamefont{N.}~\bibnamefont{Killoran}},
  \bibinfo{author}{\bibfnamefont{M.}~\bibnamefont{Cramer}}, \bibnamefont{and}
  \bibinfo{author}{\bibfnamefont{M.~B.} \bibnamefont{Plenio}},
  \bibinfo{journal}{Phys. Rev. Lett.} \textbf{\bibinfo{volume}{112}},
  \bibinfo{pages}{150501} (\bibinfo{year}{2014}).

\bibitem[{\citenamefont{Sciara et~al.}(2017)\citenamefont{Sciara, {Lo Franco},
  and Compagno}}]{sciaraSchmidt}
\bibinfo{author}{\bibfnamefont{S.}~\bibnamefont{Sciara}},
  \bibinfo{author}{\bibfnamefont{R.}~\bibnamefont{{Lo Franco}}},
  \bibnamefont{and} \bibinfo{author}{\bibfnamefont{G.}~\bibnamefont{Compagno}},
  \bibinfo{journal}{Sci. Rep.} \textbf{\bibinfo{volume}{7}},
  \bibinfo{pages}{44675} (\bibinfo{year}{2017}).

\bibitem[{\citenamefont{Lo~Franco and Compagno}(2016)}]{nolabelappr}
\bibinfo{author}{\bibfnamefont{R.}~\bibnamefont{Lo~Franco}} \bibnamefont{and}
  \bibinfo{author}{\bibfnamefont{G.}~\bibnamefont{Compagno}},
  \bibinfo{journal}{Sci. Rep.} \textbf{\bibinfo{volume}{6}},
  \bibinfo{pages}{20603} (\bibinfo{year}{2016}).

\bibitem[{\citenamefont{Compagno et~al.}(2018)\citenamefont{Compagno,
  Castellini, and Lo~Franco}}]{compagno2018dealing}
\bibinfo{author}{\bibfnamefont{G.}~\bibnamefont{Compagno}},
  \bibinfo{author}{\bibfnamefont{A.}~\bibnamefont{Castellini}},
  \bibnamefont{and}
  \bibinfo{author}{\bibfnamefont{R.}~\bibnamefont{Lo~Franco}},
  \bibinfo{journal}{Phil. Trans. R. Soc. A} \textbf{\bibinfo{volume}{376}},
  \bibinfo{pages}{20170317} (\bibinfo{year}{2018}).

\bibitem[{\citenamefont{Lo~Franco and Compagno}(2018)}]{slocc}
\bibinfo{author}{\bibfnamefont{R.}~\bibnamefont{Lo~Franco}} \bibnamefont{and}
  \bibinfo{author}{\bibfnamefont{G.}~\bibnamefont{Compagno}},
  \bibinfo{journal}{Phys. Rev. Lett.} \textbf{\bibinfo{volume}{120}},
  \bibinfo{pages}{240403} (\bibinfo{year}{2018}).

\bibitem[{\citenamefont{Morris et~al.}(2020{\natexlab{b}})\citenamefont{Morris,
  Yadin, Fadel, Zibold, Treutlein, and Adesso}}]{morrisPRX}
\bibinfo{author}{\bibfnamefont{B.}~\bibnamefont{Morris}},
  \bibinfo{author}{\bibfnamefont{B.}~\bibnamefont{Yadin}},
  \bibinfo{author}{\bibfnamefont{M.}~\bibnamefont{Fadel}},
  \bibinfo{author}{\bibfnamefont{T.}~\bibnamefont{Zibold}},
  \bibinfo{author}{\bibfnamefont{P.}~\bibnamefont{Treutlein}},
  \bibnamefont{and} \bibinfo{author}{\bibfnamefont{G.}~\bibnamefont{Adesso}},
  \bibinfo{journal}{Phys. Rev. X} \textbf{\bibinfo{volume}{10}},
  \bibinfo{pages}{041012} (\bibinfo{year}{2020}{\natexlab{b}}).

\bibitem[{\citenamefont{Sun et~al.}(2020)\citenamefont{Sun, Wang, Liu, Xu, Xu,
  Li, Guo, Castellini, Nosrati, Compagno et~al.}}]{experimentalslocc}
\bibinfo{author}{\bibfnamefont{K.}~\bibnamefont{Sun}},
  \bibinfo{author}{\bibfnamefont{Y.}~\bibnamefont{Wang}},
  \bibinfo{author}{\bibfnamefont{Z.-H.} \bibnamefont{Liu}},
  \bibinfo{author}{\bibfnamefont{X.-Y.} \bibnamefont{Xu}},
  \bibinfo{author}{\bibfnamefont{J.-S.} \bibnamefont{Xu}},
  \bibinfo{author}{\bibfnamefont{C.-F.} \bibnamefont{Li}},
  \bibinfo{author}{\bibfnamefont{G.-C.} \bibnamefont{Guo}},
  \bibinfo{author}{\bibfnamefont{A.}~\bibnamefont{Castellini}},
  \bibinfo{author}{\bibfnamefont{F.}~\bibnamefont{Nosrati}},
  \bibinfo{author}{\bibfnamefont{G.}~\bibnamefont{Compagno}},
  \bibnamefont{et~al.}, \bibinfo{journal}{Opt. Lett.}
  \textbf{\bibinfo{volume}{45}}, \bibinfo{pages}{6410} (\bibinfo{year}{2020}).

\bibitem[{\citenamefont{Lee et~al.}(2021)\citenamefont{Lee, Pramanik, Cho, Lim,
  Chin, and Kim}}]{lee2021entangling}
\bibinfo{author}{\bibfnamefont{D.}~\bibnamefont{Lee}},
  \bibinfo{author}{\bibfnamefont{T.}~\bibnamefont{Pramanik}},
  \bibinfo{author}{\bibfnamefont{Y.-W.} \bibnamefont{Cho}},
  \bibinfo{author}{\bibfnamefont{H.-T.} \bibnamefont{Lim}},
  \bibinfo{author}{\bibfnamefont{S.}~\bibnamefont{Chin}}, \bibnamefont{and}
  \bibinfo{author}{\bibfnamefont{Y.-S.} \bibnamefont{Kim}},
  \bibinfo{journal}{arXiv preprint arXiv:2104.05937}  (\bibinfo{year}{2021}).

\bibitem[{\citenamefont{Nosrati
  et~al.}(2020{\natexlab{a}})\citenamefont{Nosrati, Castellini, Compagno, and
  Lo~Franco}}]{indistentanglprotection}
\bibinfo{author}{\bibfnamefont{F.}~\bibnamefont{Nosrati}},
  \bibinfo{author}{\bibfnamefont{A.}~\bibnamefont{Castellini}},
  \bibinfo{author}{\bibfnamefont{G.}~\bibnamefont{Compagno}}, \bibnamefont{and}
  \bibinfo{author}{\bibfnamefont{R.}~\bibnamefont{Lo~Franco}},
  \bibinfo{journal}{npj Quant. Inf.} \textbf{\bibinfo{volume}{6}},
  \bibinfo{pages}{1} (\bibinfo{year}{2020}{\natexlab{a}}).

\bibitem[{\citenamefont{Nosrati
  et~al.}(2020{\natexlab{b}})\citenamefont{Nosrati, Castellini, Compagno, and
  Lo~Franco}}]{indistdynamicalprotection}
\bibinfo{author}{\bibfnamefont{F.}~\bibnamefont{Nosrati}},
  \bibinfo{author}{\bibfnamefont{A.}~\bibnamefont{Castellini}},
  \bibinfo{author}{\bibfnamefont{G.}~\bibnamefont{Compagno}}, \bibnamefont{and}
  \bibinfo{author}{\bibfnamefont{R.}~\bibnamefont{Lo~Franco}},
  \bibinfo{journal}{Phys. Rev. A} \textbf{\bibinfo{volume}{102}},
  \bibinfo{pages}{062429} (\bibinfo{year}{2020}{\natexlab{b}}).

\bibitem[{\citenamefont{Piccolini et~al.}(2021)\citenamefont{Piccolini,
  Nosrati, Compagno, Livreri, Morandotti, and Lo~Franco}}]{Piccolini_2021}
\bibinfo{author}{\bibfnamefont{M.}~\bibnamefont{Piccolini}},
  \bibinfo{author}{\bibfnamefont{F.}~\bibnamefont{Nosrati}},
  \bibinfo{author}{\bibfnamefont{G.}~\bibnamefont{Compagno}},
  \bibinfo{author}{\bibfnamefont{P.}~\bibnamefont{Livreri}},
  \bibinfo{author}{\bibfnamefont{R.}~\bibnamefont{Morandotti}},
  \bibnamefont{and}
  \bibinfo{author}{\bibfnamefont{R.}~\bibnamefont{Lo~Franco}},
  \bibinfo{journal}{Entropy} \textbf{\bibinfo{volume}{23}},
  \bibinfo{pages}{708} (\bibinfo{year}{2021}).

\bibitem[{\citenamefont{Tichy et~al.}(2011)\citenamefont{Tichy, Mintert, and
  Buchleitner}}]{tichy}
\bibinfo{author}{\bibfnamefont{M.~C.} \bibnamefont{Tichy}},
  \bibinfo{author}{\bibfnamefont{F.}~\bibnamefont{Mintert}}, \bibnamefont{and}
  \bibinfo{author}{\bibfnamefont{A.}~\bibnamefont{Buchleitner}},
  \bibinfo{journal}{J. Phys. B: At. Mol. Opt. Phys.}
  \textbf{\bibinfo{volume}{44}}, \bibinfo{pages}{192001}
  (\bibinfo{year}{2011}).

\bibitem[{\citenamefont{Wootters}(1998)}]{concurrence}
\bibinfo{author}{\bibfnamefont{W.~K.} \bibnamefont{Wootters}},
  \bibinfo{journal}{Phys. Rev. Lett.} \textbf{\bibinfo{volume}{80}},
  \bibinfo{pages}{2245–2248} (\bibinfo{year}{1998}).

\bibitem[{\citenamefont{Haikka and Maniscalco}(2010)}]{Haikka_2010}
\bibinfo{author}{\bibfnamefont{P.}~\bibnamefont{Haikka}} \bibnamefont{and}
  \bibinfo{author}{\bibfnamefont{S.}~\bibnamefont{Maniscalco}},
  \bibinfo{journal}{Phys. Rev. A} \textbf{\bibinfo{volume}{81}}
  (\bibinfo{year}{2010}), ISSN \bibinfo{issn}{1094-1622},
  \urlprefix\url{http://dx.doi.org/10.1103/PhysRevA.81.052103}.

\bibitem[{\citenamefont{Nielsen and Chuang}(2010)}]{nielsen2010quantum}
\bibinfo{author}{\bibfnamefont{M.~A.} \bibnamefont{Nielsen}} \bibnamefont{and}
  \bibinfo{author}{\bibfnamefont{I.~L.} \bibnamefont{Chuang}},
  \emph{\bibinfo{title}{Quantum Computation and Quantum Information}}
  (\bibinfo{publisher}{Cambridge University Press}, \bibinfo{year}{2010}).

\bibitem[{\citenamefont{Bellomo et~al.}(2007)\citenamefont{Bellomo, Lo~Franco,
  and Compagno}}]{Bellomo_2007}
\bibinfo{author}{\bibfnamefont{B.}~\bibnamefont{Bellomo}},
  \bibinfo{author}{\bibfnamefont{R.}~\bibnamefont{Lo~Franco}},
  \bibnamefont{and} \bibinfo{author}{\bibfnamefont{G.}~\bibnamefont{Compagno}},
  \bibinfo{journal}{Phys. Rev. Lett.} \textbf{\bibinfo{volume}{99}}
  (\bibinfo{year}{2007}).

\bibitem[{\citenamefont{Lindblad}(1976)}]{lindblad1976generators}
\bibinfo{author}{\bibfnamefont{G.}~\bibnamefont{Lindblad}},
  \bibinfo{journal}{Communications in Mathematical Physics}
  \textbf{\bibinfo{volume}{48}}, \bibinfo{pages}{119} (\bibinfo{year}{1976}).

\bibitem[{\citenamefont{Gorini et~al.}(1976)\citenamefont{Gorini, Kossakowski,
  and Sudarshan}}]{gorini1976completely}
\bibinfo{author}{\bibfnamefont{V.}~\bibnamefont{Gorini}},
  \bibinfo{author}{\bibfnamefont{A.}~\bibnamefont{Kossakowski}},
  \bibnamefont{and} \bibinfo{author}{\bibfnamefont{E.~C.~G.}
  \bibnamefont{Sudarshan}}, \bibinfo{journal}{J. Math. Phys.}
  \textbf{\bibinfo{volume}{17}}, \bibinfo{pages}{821} (\bibinfo{year}{1976}).

\bibitem[{\citenamefont{Brasil et~al.}(2013)\citenamefont{Brasil, Fanchini, and
  Napolitano}}]{brasil2013simple}
\bibinfo{author}{\bibfnamefont{C.~A.} \bibnamefont{Brasil}},
  \bibinfo{author}{\bibfnamefont{F.~F.} \bibnamefont{Fanchini}},
  \bibnamefont{and} \bibinfo{author}{\bibfnamefont{R.~d.~J.}
  \bibnamefont{Napolitano}}, \bibinfo{journal}{Revista Brasileira de Ensino de
  F{\'\i}sica} \textbf{\bibinfo{volume}{35}}, \bibinfo{pages}{01}
  (\bibinfo{year}{2013}).

\bibitem[{\citenamefont{Manzano}(2020)}]{manzano2020short}
\bibinfo{author}{\bibfnamefont{D.}~\bibnamefont{Manzano}},
  \bibinfo{journal}{AIP Advances} \textbf{\bibinfo{volume}{10}},
  \bibinfo{pages}{025106} (\bibinfo{year}{2020}).

\bibitem[{\citenamefont{Zhou et~al.}(2010)\citenamefont{Zhou, Lang, and
  Joynt}}]{zhou}
\bibinfo{author}{\bibfnamefont{D.}~\bibnamefont{Zhou}},
  \bibinfo{author}{\bibfnamefont{A.}~\bibnamefont{Lang}}, \bibnamefont{and}
  \bibinfo{author}{\bibfnamefont{R.}~\bibnamefont{Joynt}},
  \bibinfo{journal}{Quantum Information Processing}
  \textbf{\bibinfo{volume}{9}}, \bibinfo{pages}{727} (\bibinfo{year}{2010}).

\bibitem[{\citenamefont{Lo~Franco
  et~al.}(2012{\natexlab{b}})\citenamefont{Lo~Franco, D'Arrigo, Falci,
  Compagno, and Paladino}}]{lo_franco_2012}
\bibinfo{author}{\bibfnamefont{R.}~\bibnamefont{Lo~Franco}},
  \bibinfo{author}{\bibfnamefont{A.}~\bibnamefont{D'Arrigo}},
  \bibinfo{author}{\bibfnamefont{G.}~\bibnamefont{Falci}},
  \bibinfo{author}{\bibfnamefont{G.}~\bibnamefont{Compagno}}, \bibnamefont{and}
  \bibinfo{author}{\bibfnamefont{E.}~\bibnamefont{Paladino}},
  \bibinfo{journal}{Physica Scripta} \textbf{\bibinfo{volume}{2012}},
  \bibinfo{pages}{014019} (\bibinfo{year}{2012}{\natexlab{b}}).

\bibitem[{\citenamefont{Bordone et~al.}(2012)\citenamefont{Bordone, Buscemi,
  and Benedetti}}]{bordone}
\bibinfo{author}{\bibfnamefont{P.}~\bibnamefont{Bordone}},
  \bibinfo{author}{\bibfnamefont{F.}~\bibnamefont{Buscemi}}, \bibnamefont{and}
  \bibinfo{author}{\bibfnamefont{C.}~\bibnamefont{Benedetti}},
  \bibinfo{journal}{Fluctuation and Noise Letters}
  \textbf{\bibinfo{volume}{11}}, \bibinfo{pages}{1242003}
  (\bibinfo{year}{2012}).

\bibitem[{\citenamefont{Bellomo et~al.}(2012)\citenamefont{Bellomo, Lo~Franco,
  Andersson, Cresser, and Compagno}}]{bellomo_2012}
\bibinfo{author}{\bibfnamefont{B.}~\bibnamefont{Bellomo}},
  \bibinfo{author}{\bibfnamefont{R.}~\bibnamefont{Lo~Franco}},
  \bibinfo{author}{\bibfnamefont{E.}~\bibnamefont{Andersson}},
  \bibinfo{author}{\bibfnamefont{J.~D.} \bibnamefont{Cresser}},
  \bibnamefont{and} \bibinfo{author}{\bibfnamefont{G.}~\bibnamefont{Compagno}},
  \bibinfo{journal}{Phys. Scr.} \textbf{\bibinfo{volume}{2012}},
  \bibinfo{pages}{014004} (\bibinfo{year}{2012}).

\bibitem[{\citenamefont{Cai}(2020)}]{cai}
\bibinfo{author}{\bibfnamefont{X.}~\bibnamefont{Cai}}, \bibinfo{journal}{Sci.
  Rep.} \textbf{\bibinfo{volume}{10}}, \bibinfo{pages}{1}
  (\bibinfo{year}{2020}).

\bibitem[{\citenamefont{Wold et~al.}(2012)\citenamefont{Wold, Brox, Galperin,
  and Bergli}}]{wold}
\bibinfo{author}{\bibfnamefont{H.~J.} \bibnamefont{Wold}},
  \bibinfo{author}{\bibfnamefont{H.}~\bibnamefont{Brox}},
  \bibinfo{author}{\bibfnamefont{Y.~M.} \bibnamefont{Galperin}},
  \bibnamefont{and} \bibinfo{author}{\bibfnamefont{J.}~\bibnamefont{Bergli}},
  \bibinfo{journal}{Phys. Rev. B} \textbf{\bibinfo{volume}{86}},
  \bibinfo{pages}{205404} (\bibinfo{year}{2012}).

\bibitem[{\citenamefont{Kasprzak et~al.}(2006)\citenamefont{Kasprzak, Richard,
  Kundermann, Baas, Jeambrun, Keeling, Marchetti, Szyma{\'n}ska, Andr{\'e},
  Staehli et~al.}}]{kasprzak2006bose}
\bibinfo{author}{\bibfnamefont{J.}~\bibnamefont{Kasprzak}},
  \bibinfo{author}{\bibfnamefont{M.}~\bibnamefont{Richard}},
  \bibinfo{author}{\bibfnamefont{S.}~\bibnamefont{Kundermann}},
  \bibinfo{author}{\bibfnamefont{A.}~\bibnamefont{Baas}},
  \bibinfo{author}{\bibfnamefont{P.}~\bibnamefont{Jeambrun}},
  \bibinfo{author}{\bibfnamefont{J.}~\bibnamefont{Keeling}},
  \bibinfo{author}{\bibfnamefont{F.}~\bibnamefont{Marchetti}},
  \bibinfo{author}{\bibfnamefont{M.}~\bibnamefont{Szyma{\'n}ska}},
  \bibinfo{author}{\bibfnamefont{R.}~\bibnamefont{Andr{\'e}}},
  \bibinfo{author}{\bibfnamefont{J.}~\bibnamefont{Staehli}},
  \bibnamefont{et~al.}, \bibinfo{journal}{Nature}
  \textbf{\bibinfo{volume}{443}}, \bibinfo{pages}{409} (\bibinfo{year}{2006}).

\bibitem[{\citenamefont{Zipkes et~al.}(2010)\citenamefont{Zipkes, Palzer, Sias,
  and K{\"o}hl}}]{zipkes2010trapped}
\bibinfo{author}{\bibfnamefont{C.}~\bibnamefont{Zipkes}},
  \bibinfo{author}{\bibfnamefont{S.}~\bibnamefont{Palzer}},
  \bibinfo{author}{\bibfnamefont{C.}~\bibnamefont{Sias}}, \bibnamefont{and}
  \bibinfo{author}{\bibfnamefont{M.}~\bibnamefont{K{\"o}hl}},
  \bibinfo{journal}{Nature} \textbf{\bibinfo{volume}{464}},
  \bibinfo{pages}{388} (\bibinfo{year}{2010}).

\bibitem[{\citenamefont{Puentes et~al.}(2005)\citenamefont{Puentes, Voigt,
  Aiello, and Woerdman}}]{puentes2005experimental}
\bibinfo{author}{\bibfnamefont{G.}~\bibnamefont{Puentes}},
  \bibinfo{author}{\bibfnamefont{D.}~\bibnamefont{Voigt}},
  \bibinfo{author}{\bibfnamefont{A.}~\bibnamefont{Aiello}}, \bibnamefont{and}
  \bibinfo{author}{\bibfnamefont{J.}~\bibnamefont{Woerdman}},
  \bibinfo{journal}{Opt. Lett.} \textbf{\bibinfo{volume}{30}},
  \bibinfo{pages}{3216} (\bibinfo{year}{2005}).

\bibitem[{\citenamefont{Puentes et~al.}(2007)\citenamefont{Puentes, Aiello,
  Voigt, and Woerdman}}]{puentes2007entangled}
\bibinfo{author}{\bibfnamefont{G.}~\bibnamefont{Puentes}},
  \bibinfo{author}{\bibfnamefont{A.}~\bibnamefont{Aiello}},
  \bibinfo{author}{\bibfnamefont{D.}~\bibnamefont{Voigt}}, \bibnamefont{and}
  \bibinfo{author}{\bibfnamefont{J.}~\bibnamefont{Woerdman}},
  \bibinfo{journal}{Phys. Rev. A} \textbf{\bibinfo{volume}{75}},
  \bibinfo{pages}{032319} (\bibinfo{year}{2007}).

\bibitem[{\citenamefont{Shaham and Eisenberg}(2011)}]{shaham2011realizing}
\bibinfo{author}{\bibfnamefont{A.}~\bibnamefont{Shaham}} \bibnamefont{and}
  \bibinfo{author}{\bibfnamefont{H.}~\bibnamefont{Eisenberg}},
  \bibinfo{journal}{Phys. Rev. A} \textbf{\bibinfo{volume}{83}},
  \bibinfo{pages}{022303} (\bibinfo{year}{2011}).

\bibitem[{\citenamefont{Xin et~al.}(2017)\citenamefont{Xin, Wei, Pedernales,
  Solano, and Long}}]{xin2017quantum}
\bibinfo{author}{\bibfnamefont{T.}~\bibnamefont{Xin}},
  \bibinfo{author}{\bibfnamefont{S.-J.} \bibnamefont{Wei}},
  \bibinfo{author}{\bibfnamefont{J.~S.} \bibnamefont{Pedernales}},
  \bibinfo{author}{\bibfnamefont{E.}~\bibnamefont{Solano}}, \bibnamefont{and}
  \bibinfo{author}{\bibfnamefont{G.-L.} \bibnamefont{Long}},
  \bibinfo{journal}{Phys. Rev. A} \textbf{\bibinfo{volume}{96}},
  \bibinfo{pages}{062303} (\bibinfo{year}{2017}).

\bibitem[{\citenamefont{Ryan et~al.}(2009)\citenamefont{Ryan, Laforest, and
  Laflamme}}]{ryan2009randomized}
\bibinfo{author}{\bibfnamefont{C.}~\bibnamefont{Ryan}},
  \bibinfo{author}{\bibfnamefont{M.}~\bibnamefont{Laforest}}, \bibnamefont{and}
  \bibinfo{author}{\bibfnamefont{R.}~\bibnamefont{Laflamme}},
  \bibinfo{journal}{New J. Phys.} \textbf{\bibinfo{volume}{11}},
  \bibinfo{pages}{013034} (\bibinfo{year}{2009}).

\bibitem[{\citenamefont{Bruzewicz et~al.}(2019)\citenamefont{Bruzewicz,
  Chiaverini, McConnell, and Sage}}]{bruzewicz}
\bibinfo{author}{\bibfnamefont{C.}~\bibnamefont{Bruzewicz}},
  \bibinfo{author}{\bibfnamefont{J.}~\bibnamefont{Chiaverini}},
  \bibinfo{author}{\bibfnamefont{R.}~\bibnamefont{McConnell}},
  \bibnamefont{and} \bibinfo{author}{\bibfnamefont{J.}~\bibnamefont{Sage}},
  \bibinfo{journal}{Appl. Phys. Rev.} \textbf{\bibinfo{volume}{6}},
  \bibinfo{pages}{021314} (\bibinfo{year}{2019}).

\bibitem[{\citenamefont{Myatt et~al.}(2000)\citenamefont{Myatt, King,
  Turchette, Sackett, Kielpinski, Itano, Monroe, and Wineland}}]{myatt}
\bibinfo{author}{\bibfnamefont{C.~J.} \bibnamefont{Myatt}},
  \bibinfo{author}{\bibfnamefont{B.~E.} \bibnamefont{King}},
  \bibinfo{author}{\bibfnamefont{Q.~A.} \bibnamefont{Turchette}},
  \bibinfo{author}{\bibfnamefont{C.~A.} \bibnamefont{Sackett}},
  \bibinfo{author}{\bibfnamefont{D.}~\bibnamefont{Kielpinski}},
  \bibinfo{author}{\bibfnamefont{W.~M.} \bibnamefont{Itano}},
  \bibinfo{author}{\bibfnamefont{C.}~\bibnamefont{Monroe}}, \bibnamefont{and}
  \bibinfo{author}{\bibfnamefont{D.~J.} \bibnamefont{Wineland}},
  \bibinfo{journal}{Nature} \textbf{\bibinfo{volume}{403}},
  \bibinfo{pages}{269} (\bibinfo{year}{2000}).

\bibitem[{\citenamefont{Schindler et~al.}(2013)\citenamefont{Schindler, Nigg,
  Monz, Barreiro, Martinez, Wang, Quint, Brandl, Nebendahl, Roos
  et~al.}}]{schindler}
\bibinfo{author}{\bibfnamefont{P.}~\bibnamefont{Schindler}},
  \bibinfo{author}{\bibfnamefont{D.}~\bibnamefont{Nigg}},
  \bibinfo{author}{\bibfnamefont{T.}~\bibnamefont{Monz}},
  \bibinfo{author}{\bibfnamefont{J.~T.} \bibnamefont{Barreiro}},
  \bibinfo{author}{\bibfnamefont{E.}~\bibnamefont{Martinez}},
  \bibinfo{author}{\bibfnamefont{S.~X.} \bibnamefont{Wang}},
  \bibinfo{author}{\bibfnamefont{S.}~\bibnamefont{Quint}},
  \bibinfo{author}{\bibfnamefont{M.~F.} \bibnamefont{Brandl}},
  \bibinfo{author}{\bibfnamefont{V.}~\bibnamefont{Nebendahl}},
  \bibinfo{author}{\bibfnamefont{C.~F.} \bibnamefont{Roos}},
  \bibnamefont{et~al.}, \bibinfo{journal}{New J. Phys.}
  \textbf{\bibinfo{volume}{15}}, \bibinfo{pages}{123012}
  (\bibinfo{year}{2013}).

\bibitem[{\citenamefont{Blais et~al.}(2007)\citenamefont{Blais, Gambetta,
  Wallraff, Schuster, Girvin, Devoret, and Schoelkopf}}]{blais_2007}
\bibinfo{author}{\bibfnamefont{A.}~\bibnamefont{Blais}},
  \bibinfo{author}{\bibfnamefont{J.}~\bibnamefont{Gambetta}},
  \bibinfo{author}{\bibfnamefont{A.}~\bibnamefont{Wallraff}},
  \bibinfo{author}{\bibfnamefont{D.~I.} \bibnamefont{Schuster}},
  \bibinfo{author}{\bibfnamefont{S.~M.} \bibnamefont{Girvin}},
  \bibinfo{author}{\bibfnamefont{M.~H.} \bibnamefont{Devoret}},
  \bibnamefont{and} \bibinfo{author}{\bibfnamefont{R.~J.}
  \bibnamefont{Schoelkopf}}, \bibinfo{journal}{Phys. Rev. A}
  \textbf{\bibinfo{volume}{75}}, \bibinfo{pages}{032329}
  (\bibinfo{year}{2007}).

\bibitem[{\citenamefont{Blais et~al.}(2020)\citenamefont{Blais, Girvin, and
  Oliver}}]{blais_2020}
\bibinfo{author}{\bibfnamefont{A.}~\bibnamefont{Blais}},
  \bibinfo{author}{\bibfnamefont{S.~M.} \bibnamefont{Girvin}},
  \bibnamefont{and} \bibinfo{author}{\bibfnamefont{W.~D.}
  \bibnamefont{Oliver}}, \bibinfo{journal}{Nat. Phys.}
  \textbf{\bibinfo{volume}{16}}, \bibinfo{pages}{247} (\bibinfo{year}{2020}).

\bibitem[{\citenamefont{Giovannetti and Fazio}(2005)}]{giovannetti}
\bibinfo{author}{\bibfnamefont{V.}~\bibnamefont{Giovannetti}} \bibnamefont{and}
  \bibinfo{author}{\bibfnamefont{R.}~\bibnamefont{Fazio}},
  \bibinfo{journal}{Phys. Rev. A} \textbf{\bibinfo{volume}{71}},
  \bibinfo{pages}{032314} (\bibinfo{year}{2005}).

\end{thebibliography}

\end{document}